\documentclass[usenatbib,useAMS,letterpaper]{mnras}
\usepackage{float}
\usepackage{graphicx}
\usepackage{epstopdf}
\usepackage{ textcomp }
\usepackage{multirow}
\usepackage{ctable}
\usepackage{url}
\usepackage{times}
\usepackage{amsmath}
\usepackage{amssymb}
\usepackage{xspace}
\usepackage{mathrsfs} 
\usepackage{tabularx}
\usepackage{fixltx2e}
\usepackage{color}
\usepackage{placeins}
\usepackage{comment}

\newcommand{\acknowledgments}{\begin{small}\section*{Acknowledgements}\end{small}}

\defcitealias{2019MNRAS.487.4393S}{Paper {\small I}}
\defcitealias{2020MNRAS.491.1190S}{Paper {\small II}}
\newcommand\sref[1]{\hyperref[#1]{\S~\ref*{#1}}}
\newcommand\fref[1]{\hyperref[#1]{Fig.~\ref*{#1}}}
\newcommand\Eqref[1]{Eq.~(\hyperref[#1]{\ref*{#1}})}
\newcommand\eeqref[1]{Eq.~\hyperref[#1]{\ref*{#1}}}

\newcommand\tref[1]{\hyperref[#1]{Table~\ref*{#1}}}
\newcommand\aref[1]{\hyperref[#1]{Appendix~\ref*{#1}}}

\newcommand{\ksu}[1]{\textcolor{black}{#1}}
\newcommand{\ksuu}[1]{\textcolor{black}{#1}}

\newcommand{\glb}[1]{\textcolor{black}{#1}}
\newcommand{\zh}[1]{\textcolor{black}{#1}}

\newcommand{\hlt}[1]{\textcolor{blue}{#1}}

\newcommand{\oneline}[1]{%
  \newdimen{\namewidth}%
  \setlength{\namewidth}{\widthof{#1}}%
  \ifthenelse{\lengthtest{\namewidth < \textwidth}}%
  {#1}
  {\resizebox{\textwidth}{!}{#1}}
}
 
\title[Self-regulation of early BH accretion via jets]{\zh{Self-regulation of black hole accretion via jets in early protogalaxies}}
\author[]{
\parbox[t]{\textwidth}{
Kung-Yi Su$^{1,2}$\thanks{E-mail: k.su@columbia.edu},  Greg L. Bryan$^{1}$, Zolt\'{a}n Haiman$^{1}$,  Rachel S. Somerville$^{2}$, Christopher C. Hayward$^{2}$,  Claude-Andr\'{e} Faucher-Gigu\`{e}re$^3$}
\vspace*{6pt} \\
$^1$Department of Astronomy, Columbia University, 550 West 120th Street, New York, NY 10027, USA\\
$^2$Center for Computational Astrophysics, Flatiron Institute, 162 Fifth Avenue, New York, NY 10010, USA\\
$^3$Department of Physics \& Astronomy and CIERA, Northwestern University, 1800 Sherman Ave, Evanston, IL 60201, USA
}
\usepackage[all]{hypcap}

\begin{document}
\long\def\/*#1*/{}
\date{Submitted to MNRAS}

\pagerange{\pageref{firstpage}--\pageref{lastpage}} \pubyear{2021}

\maketitle

\label{firstpage}

\begin{abstract}
\zh{The early growth of black holes in high-redshift galaxies is likely regulated by their feedback on the surrounding gas. While radiative feedback has been extensively studied, the role of mechanical feedback has received comparatively less scrutiny to date.   Here} we use high-resolution parsec-scale hydrodynamical simulations to study jet propagation and its effect on black hole accretion onto 100 ${\rm M_\odot}$ black holes in \zh{the} dense, low-metallicity gas \zh{expected in early protogalaxies}. As the jet propagates, it shocks the surrounding gas and forms a jet cocoon. The cocoon consists of a rapidly-cooling cold phase at the interface with the background gas and an over-pressured subsonic phase of reverse shock-heated gas filling the cocoon interior. \zh{We systematically vary the background gas density and temperature, black hole feedback efficiency, and the jet model}. We found that the width of the jet cocoon roughly follows a scaling derived by assuming momentum conservation in the jet propagation direction, and energy conservation in the lateral directions. Depending on the assumed gas and jet properties, the cocoon can either stay elongated out to a large radius or isotropize before reaching the Bondi radius, forming a nearly spherical bubble. Lower jet velocities and higher background gas densities result in self-regulation to higher momentum fluxes and elongated cocoons. In all cases, the 
outward momentum flux of the
cocoon balances the inward momentum flux of the inflowing gas near the Bondi radius, which ultimately regulates black hole accretion.\ksu{We also examine the accretion variability and find that the larger the distance the jet cocoon reaches (either due to lower temperature or a more elongated jet cocoon), the longer the variability timescale of the black hole accretion rate.} 
\zh{Overall, we find that the time-averaged accretion rate always remains below the Bondi rate, and exceeds the Eddington rate only if the ambient medium is dense and cold, and/or the jet is weak (low velocity and mass-loading).  We derive the combination of jet and ambient gas parameters yielding super-Eddington growth.}
\end{abstract}


\section{Introduction}
\label{S:intro}

\zh{
The origin of supermassive black holes (SMBHs) with masses of $\gtrsim 10^9~{\rm M_\odot}$, powering bright quasars observed in the first billion years after the Big Bang (redshifts $z\gtrsim 6$; see, e.g.~\citealt{Bosman2022} for an up-to-date compilation) remains an unsolved puzzle.  Proposed explanations range from rapid, super-Eddington growth of stellar mass seed black holes (BHs), the ``direct collapse" of a supermassive star, to runaway mergers between stellar-mass objects, as well as more exotic phenomena~\citep[see, e.g.][for recent comprehensive reviews]{2020ARA&A..58...27I,Volonteri+2021}.}

\zh{One promising scenario is for a low-mass seed BH to grow at rates well above the fiducial Eddington rate $\dot{M}_{\rm Edd}\equiv L_{\rm Edd}/\epsilon c^2$ (where $L_{\rm Edd}$ is the Eddington luminosity, $c$ is the speed of light, and $\epsilon$ is a radiative efficiency).  Indeed, small-scale simulations of BH accretion show that BHs surrounded by dense gas can accrete at rates up to at least $\sim 100~\dot{M}_{\rm Edd}$~\citep[e.g.,][]{Jiang+2014,Sadowski+2014}.
However, feedback from the BH accretion itself poses possible obstacles to sustaining such rapid growth. Even in the presence of dense ambient gas, allowing rapid fueling, radiative feedback on large scales tends to make the accretion episodic, with a strongly suppressed time-averaged accreton rate~\citep[e.g.][]{Milos+2009,2011ApJ...739....2P}. BH radiation may also outright eject gas from the shallow gravitational potential of its low-mass parent halo, preventing rapid accretion~\citep{AlvarezAbel2009}.   On the other hand, these deleterious radiative effects may be avoided in the hyper-Eddington regime, in which radiation is trapped and cannot exert large-scale feedback~\citep{Inayoshi+2016,2020MNRAS.497..302T}.}

\zh{In addition to radiative feedback, mechanical feedback presents another potential obstacle to rapid and sustained BH growth.  While such mechanical feedback has been less explored in the high-redshift context, it is well established to play a crucial role in galaxy formation and evolution at lower redshifts.}
Active galactic nucleus (AGN) feedback is known to quench star-formation in massive galaxies and clusters, keeping them ``red and dead'' over a significant fraction of cosmic time.
Among the different forms of AGN feedback, extensive galaxy-scale simulations have shown that AGN jet models are, in principle, capable of quenching a galaxy and stopping the cooling flows \citep[e.g.,][]{2010MNRAS.409..985D,2012MNRAS.424..190G,2012MNRAS.427.1614Y,2014ApJ...789...54L,2015ApJ...811...73L,2015ApJ...811..108P,2016ApJ...818..181Y,2017ApJ...834..208R,2017MNRAS.472.4707B,2019MNRAS.483.2465M,2020MNRAS.491.1190S}. 
 Observational studies also infer that AGN can provide an energy budget comparable to the cooling rate \citep{2004ApJ...607..800B}. There are also observations of unambiguous cases of AGN expelling gas from galaxies, injecting thermal energy via shocks or sound waves, via photo-ionization and Compton heating, or via ``stirring'' the circum-galactic medium (CGM) and intra-cluster medium (ICM). This can create ``bubbles'' of hot plasma with non-negligible relativistic components, which are ubiquitous around massive galaxies \citep[see, e.g.,][for a detailed review]{2012ARA&A..50..455F,2018ARA&A..56..625H}. In \cite{2021MNRAS.507..175S} and Su et al. (in prep.), we carried out a broad parameter study of AGN jets in $10^{12}-10^{15} {\rm M_\odot}$ clusters and found a subset of models which inflate a sufficiently large cocoon with a long enough cooling time that these jets can quench the \zh{central} galaxy. 
 
In addition to the thoroughly studied cases of SMBHs in massive galaxies, various studies also suggested AGN feedback in much smaller dwarf galaxies and from intermediate-mass black holes ($M_{\rm BH}\sim10^2-10^5 {\rm M_\odot}$; \citealt[e.g.,][]{2017ApJ...845...50N,2018ApJ...861...50B,2018MNRAS.476..979P,2019ApJ...884..180D,2019ApJ...884...54M}), 
some of which are observed in the form of AGN jets \citep[e.g.,][]{2006ApJ...636...56G,2006ApJ...646L..95W,2008ApJ...686..838W,2011AN....332..379M,2012ApJ...753..103N,2012ApJ...750L..24R,2012Sci...337..554W,2013MNRAS.436.1546M,2013MNRAS.436.3128M,2014ApJ...787L..30R,2015MNRAS.448.1893M,2018MNRAS.478.2576M,2018MNRAS.480L..74M,2019MNRAS.488..685M}. Unsurprisingly, AGN feedback can also affect the growth of these smaller black holes, alter the surrounding gas properties, and play a significant role in \zh{sculping} the galaxy they live in, especially in dwarfs and high-redshift galaxies \citep{2022arXiv220306201W}. 
 
 Observations also find \ksu{supermassive black holes ($M_{\rm BH}\gg10^{5} {\rm M_\odot}$)} at high-redshift ($z\gtrsim 4$) with jetted AGN quasars \citep[e.g.,][]{2021A&A...655A..95S,2022arXiv220309527S}. It is unclear whether a $\sim$100 ${\rm M_\odot}$ black hole, which can be presumed to produce jets, as well, if it is fed at super-Eddington rates, could sustain rapid growth onto a supermassive black hole. 
Recent work has addressed this problem in  slightly different contexts, either 
investigating the impact of 
\zh{wider-angle outflows} produced at larger radii in the accretion flow \citep[e.g.,][]{2020MNRAS.497..302T}, or by utilizing galaxy-scale simulations to assess the growth of larger black holes ($>10^4 {\rm M_\odot}$) with a jet \citep[e.g.,][]{2019MNRAS.486.3892R,2022arXiv220108766M}.  The present work aims to study how AGN jets affect accretion onto $100 {\rm M_\odot}$  ``seed" black holes in dense, low-metallicity gas, mimicking conditions expected in high-redshift protogalaxies. Additionally, we study \zh{in detail the physics of how jet-inflated cocoons propagate to large radii and self-regulate BH accretion, using analytic models to interpret our simulation results}.

In 
galaxy-scale simulations, including in our own previous work \citep[e.g,][]{2020MNRAS.497.5292T,2020MNRAS.491.1190S,2021MNRAS.507..175S,2022arXiv220306201W},  both AGN feedback and
BH accretion have been implemented with sub-grid prescriptions.  Models based on Bondi-Hoyle accretion \citep{1952MNRAS.112..195B,2005MNRAS.361..776S} and accretion via gravitational torques \citep{2011MNRAS.415.1027H,2017MNRAS.464.2840A} involve assumptions about gas properties, which might not always be valid, especially in the presence of an unresolved jet. To better address this question, in this work, we model a cloud of gas with systematically varied properties around the black hole at sufficiently high resolution (with the minimum gravitational force softening at least 1000 times less than the Bondi radius), to resolve the gravitational capture of individual gas particles \citep{2016MNRAS.458..816H,2021ApJ...917...53A}. We also implement various jet models to study how they affect BH accretion and how the jet propagates to a larger radius. \ksu{Although the jets in this study are launched at a much smaller scale, the initial jet itself remains sub-grid relative to the scales we can resolve. This work also addresses how sub-grid jet models launched on different scales connect to each other. 
We also parameterize the results of our simulations in order to provide the effective \zh{long-term time-averaged} accretion rate, given different gas properties beyond the Bondi radius and with different jet models. }. \zh{We delineate the parameter space of gas and jet properties over which super-Eddington growth may occur.}

The rest of this paper is organised as follows.
In \sref{S:methods}, we summarise our initial conditions (ICs), black hole accretion model and the AGN jet parameters we survey, and we describe our numerical simulations. We present the results with different jet velocities, which show the most dramatic differences, in \sref{S:j_vel_difference}. We develop a toy model describing the regulation of different jet models in different environments in \sref{S:toy}. We present \zh{a suite of additional simulations with varying model parameters} and compare the results with the toy model in \sref{s:result}. We compare our study to several other recent works, and summarise the implications of our findings in \sref{S:discussion}. We enumerate our main conclusions in \sref{S:conclusions}. 
We include a set of simulations exploring numerical choices, as well as resolution studies, in \aref{a:res}.

\section{Methodology} \label{S:methods}
We perform simulations of a box of gas under the effect of jet feedback from a
 100 ${\rm M_\odot}$
 black hole. Our simulations use {\sc GIZMO}\footnote{A public version of this code is available at \href{http://www.tapir.caltech.edu/~phopkins/Site/GIZMO.html}{\textit{http://www.tapir.caltech.edu/$\sim$phopkins/Site/GIZMO.html}}}  \citep{2015MNRAS.450...53H}, in its meshless finite mass (MFM) mode, which is a Lagrangian mesh-free Godunov method, capturing advantages of both grid-based and smoothed-particle hydrodynamics (SPH) methods. Numerical implementation details and extensive tests are presented in a series of methods papers for, e.g.,\ hydrodynamics and self-gravity \citep{2015MNRAS.450...53H}. All of our simulations employ the FIRE-2 implementation of cooling (followed from $10-10^{10}$K), including the effects of photo-electric and photo-ionization heating, collisional, Compton, fine-structure, recombination, atomic, and molecular cooling \citep[following][]{2017arXiv170206148H}. Note that we impose a temperature floor at $T_\infty$, which will be specified in the initial conditions (and systematically varied), assuming other feedback processes not included in these simulations keep the gas from cooling further. 
\glb{We assume a metallicity of $10^{-4}$ $Z_{\odot}$, \zh{which may be expected in the protogalaxies hosting the first stellar-mass BH seeds, and} which is sufficiently low that metal cooling above $10^3$ K (the lowest value of $T_\infty$ that we adopt) is negligible.}


\subsection{Initial conditions}
Ideally, we would evolve the black hole accretion within the context of a cosmological simulation that resolves the gas dynamics at high redshift \citep[e.g., as done for minihalos by][]{2009ApJ...701L.133A}. However, given the very large uncertainties in high-redshift conditions, we instead approximate the physical conditions near the black hole as a uniform patch of gas. This allows us to systematically vary the gas properties in order to understand how these impact the black hole regulation. In particular, the initial condition we adopt is a uniform 3D-box of uniformly-distributed gas particles with constant density and temperature, which we denote $n_\infty$ and $T_\infty$. A 100 ${\rm M_\odot}$ black hole is placed at the center of the box.  \zh{As mentioned above,} the initial metallicity of the gas is set to a very low value ($10^{-4}$ $Z_{\odot}$).  

To achieve a higher resolution in the vicinity of the black hole, 
\zh{where the accretion occurs and the overall regulation is determined,} and in the vicinity of the jet,
we use a hierarchical super-Lagrangian refinement scheme \citep{2020MNRAS.491.1190S,2021MNRAS.507..175S} to reach $\sim 1.4 \times 10^{-6}\,{\rm M}_{\sun} (n_\infty/10^4 {\rm cm}^{-3})$ mass resolution around the z-axis where the jet is launched, much higher than many previous global studies \citep[e.g.,][]{2017MNRAS.470.4530W,2021MNRAS.507..175S}.
The mass resolution decreases as a function of distance from the z-axis ($r_{\rm 2d}$), roughly proportional to  $r_{\rm 2d}$ for $r_{\rm 2d}>2.5\times10^{-2}$ pc. The numerical details are summarized in \tref{tab:run}. The highest resolution region is where  $r_{\rm 2d}$ is smaller than $r_{\rm 2d}=2.5\times10^{-3}$ pc unless otherwise stated. A resolution study is presented in \aref{a:res}.

\subsection{BH accretion}

As discussed in the introduction, black hole accretion is not modelled with the Bondi assumption, but instead is determined by following the gravitational capture of gas \citep{2016MNRAS.458..816H,2021ApJ...917...53A} directly, \zh{and implementing its subsequent accretion onto the black hole via an $\alpha$-disk prescription (see below)}. A gas particle is accreted if it is gravitationally bound to the black hole and the estimated apocentric radius is smaller than $r_{\rm acc}$\footnote{This provides a  scale for the sub-grid accretion model and for the $\alpha$-disk model.}. This sink radius $r_{\rm acc}$ is set to $3\times10^{-5} - 1.5\times10^{-4}$ pc according to the black hole neighborhood gas density. In more detail, \ksu{the sink radius $r_{\rm acc}$ is set to be a radius from the black hole enclosing 96 ``weighted'' neighborhood gas particles but capped to be within  ($3\times10^{-5}-1.5\times10^{-4}$ pc).}

Although we follow the gas down to distances very close to the black hole, we do not model the accretion disk itself, but instead adopt a simple $\alpha$-disk model. The accreted gas adds to the $\alpha$-disk mass ($M_{\rm \alpha}$, which is initially set to zero). The mass in the $\alpha$-disk is then supplied to the black hole at the rate
\begin{align}
\dot{M}_{\rm acc}= M_{\rm \alpha}/t_{\rm disk}.
\end{align}
We assume a constant $t_{\rm disk}=1000$ year from an estimated viscous time scale of a \cite{1973A&A....24..337S} disk, assuming the accretion disk is at $10^4 K$, as $t_{\rm disk}\sim t_{\rm ff} \mathcal{M}^2/\alpha\sim 1000 {\rm yr} \left(M_{\rm BH}/100 {\rm M_\odot}\right)^{1/2} \left(\alpha/0.1\right)^{-1} \left(r_{\rm acc}/10^{-4} \rm pc\right)^{1/2}$,
where $t_{\rm ff}$ is the free fall time at $r_{\rm acc}$ and  $\mathcal{M}$ is the Mach number of gas in the $\alpha$-disk. In \aref{a:res} we explore the impact of varying $t_{\rm disk}$.

\subsection{Jet models}
We adopt a jet model following \cite{2021MNRAS.507..175S}. 
In brief, a jet is launched with a particle-spawning method, which creates new gas cells ("resolution elements") to represent the jet material.
The spawned particles have a fixed initial
mass, temperature, and velocity, which sets the specific energy of the jet.
With this method we have better control of the jet properties, as launching using particle spawning depends less on local gas properties than when depositing energy/momentum based on the distribution of neighboring gas elements \footnote{\ksu{The traditional method usually does a particle neighbor search from the black hole and dumps the designated energy and momentum into these gas particles. Therefore, the effect will depend on the local gas properties and the exact geometric distribution. See \cite{2022arXiv220306201W} for a comparison of different methods.}}. We can also enforce a higher resolution for the jet elements, allowing light jets to be accurately modeled.  
The spawned gas particles have a mass resolution as indicated in \tref{tab:run} and are forbidden to de-refine (merge into a common gas element) before they decelerate to 10\% of the launch velocity. Two particles are spawned in opposite z-directions at the same time when the accumulated jet mass flux reaches twice the target spawned particle mass, so linear momentum is always exactly conserved. Initially, the spawned particle is randomly placed on a sphere with a radius of $r_0$, which is either $10^{-5}$ pc or half the distance between the black hole and the closest gas particle, whichever is smaller. If the particle is initialized at a position $(r_0,\theta_0,\phi_0)$ \zh{in spherical polar coordinates}, and the jet opening-angle of a specific model is $\theta_{\rm op}$ \zh{(say $=1^o$, which is the case for our jet model)}, the \zh{polar angle of the} initial velocity direction of the jet will be set at $\zh{\theta_v=}2\theta_{\rm op}\theta_0/\pi$.
With this, the projected paths of any two particles will not intersect. 

We parameterize the jet mass flux with a constant feedback mass fraction
\begin{align}
\dot{M}_{\rm jet}=\eta_{\rm m, fb}\dot{M}_{\rm BH},
\end{align}
so the feedback energy and momentum fluxes are
\begin{align}
\dot{E}_{\rm jet}&=\eta_{\rm m, fb}\dot{M}_{\rm BH}  \left(\frac{1}{2} V_{\rm jet}^2+ \frac{3kT}{2\mu}  \right),\notag\\
\dot{P}_{\rm jet}&=\eta_{\rm m, fb}\dot{M}_{\rm BH}  V_{\rm jet} ,
\end{align}
where $V_{\rm jet}$ is the adopted jet velocity  and $\mu$ is the mean particle mass.

\begin{table*}
\begin{center}
 \caption{Physics variations of all simulations}
 \label{tab:run}
\resizebox{17.7cm}{!}
{
\begin{tabular}{c|cccc|ccc|cc|cccc}
\hline
\hline
&\multicolumn{4}{c|}{Numerical details} & \multicolumn{3}{c|}{Feedback parameters} & \multicolumn{2}{c|}{Background gas} & \multicolumn{4}{c}{Resulting averaged accretion rate and fluxes}\\
\hline
Model           & $\Delta T$ & Box size & $m^{\rm max}_{\rm g}$ & $m_{\rm jet}$  & $\eta_{\rm m, fb}$ & $V_{\rm jet}$ & $T_{\rm jet}$ & $n_\infty$ & $T_\infty$ & $\langle \dot{M}_{\rm BH} \rangle$ & $\frac{\langle \dot{M}_{\rm BH} \rangle}{\dot{M}_{\rm Bondi}}$ &  $\frac{\langle \dot{M}_{\rm BH} \rangle}{\dot{M}_{\rm Edd}}$  &  $\frac{\langle \dot{E}_{\rm jet} \rangle}{\dot{E}_{\rm Edd}}$ \\
                & kyr        & pc       & ${\rm M_\odot}$             & ${\rm M_\odot}$      &                    & km s$^{-1}$   & K             & cm$^{-3}$  & K & $M_\odot {\rm yr}^{-1}$ \\ 
\hline
\multicolumn{1}{c|}{\bf \hlt{Fiducial}}& \multicolumn{4}{c|}{}& \multicolumn{3}{c|}{}&\multicolumn{2}{c|}{}\\
$\eta$5e-2--$v$j1e4--n1e5--T1e4 (I,II,III)$^*$ & 100 & 0.4 & 1.4e-6 & 1e-7 &0.05 &1e4 &1e4 & 1e5 &6e3 &0.7-1.5e-7&2.1-4.6e-3 &0.031-0.067 &0.87-1.9e-5\\
\hline 
\multicolumn{1}{c|}{\bf \hlt{Feedback mass fraction}}& \multicolumn{4}{c|}{}& \multicolumn{3}{c|}{}&\multicolumn{2}{c|}{}\\
$\eta$5e-3--$v$j1e5--n1e5--T1e4 & 100 & 0.4 & 1.4e-6 & 1e-7 &0.005 &1e4 &1e4 & 1e5 &6e3 &9.4e-7&2.9e-2 &0.42    &1.2e-5\\
$\eta$5e-1--$v$j1e5--n1e5--T1e4 & 100 & 0.4 & 1.4e-6 & 1e-7 &0.5 &1e4 &1e4 & 1e5 &6e3 &1.3e-8  &4e-4 &5.8e-3 &1.6e-5\\
\hline 
\multicolumn{1}{c|}{\bf \hlt{Jet velocity}}& \multicolumn{4}{c|}{}& \multicolumn{3}{c|}{}&\multicolumn{2}{c|}{}\\
$\eta$5e-2--$v$j3e3--n1e5--T1e4 & 90 & 0.8 & 1.4e-6 & 1e-7 &0.05 &3e3 &1e4 & 1e5 &6e3 &1e-5  &0.31 &4.5 & 1.1e-4\\
$\eta$5e-2--$v$j3e4--n1e5--T1e4 & 100 & 0.4 & 1.4e-6 & 1e-7 &0.05 &3e4 &1e4 & 1e5 &6e3 &8.9e-9 &2.7e-4 &4.0e-3 &1e-5 \\
\hline
\multicolumn{1}{c|}{\bf \hlt{Thermal jet}}& \multicolumn{4}{c|}{}& \multicolumn{3}{c|}{}&\multicolumn{2}{c|}{}\\
$\eta$5e-2--Tj3e9--n1e5--T1e4 & 100 & 0.4 & 1.4e-6 & 1e-7 &0.05 &1e4 &3e9 & 1e5 &6e3 &2.4e-8 &7.4e-4 &0.011 &3.0e-6\\
\hline
\multicolumn{1}{c|}{\bf \hlt{Gas density}}& \multicolumn{4}{c|}{}& \multicolumn{3}{c|}{}&\multicolumn{2}{c|}{}\\
$\eta$5e-2--$v$j3e3--n1e2--T1e4 & 100 & 0.4 & 1.4e-9 & 1e-10 &0.05 &1e4 &1e4 & 1e2 &6e3 &1.6e-11 &4.9e-4 &7.2e-6 &2e-9\\
$\eta$5e-2--$v$j3e3--n1e3--T1e4 & 100 & 0.4 & 1.4e-8 & 1e-9 &0.05 &1e4 &1e4 & 1e3 &6e3  &2.3e-10 &7.0e-4 &1.0e-4 &2.9e-8\\
$\eta$5e-2--$v$j3e3--n1e4--T1e4 & 100 & 0.4 & 1.4e-7 & 1e-8 &0.05 &1e4 &1e4 & 1e4 &6e3 & 3.5e-9  &1.1e-3 &1.6e-3 &4.4e-7\\
$\eta$5e-2--$v$j3e3--n1e6--T1e4 & 40 & 0.8 & 1.4e-5 & 1e-6 &0.05 &1e4 &1e4 & 1e6 &6e3 &1.3e-5    &4e-2 &5.8 &1.6e-3\\
\hline
\multicolumn{1}{c|}{\bf \hlt{Gas temperature}}& \multicolumn{4}{c|}{}& \multicolumn{3}{c|}{}&\multicolumn{2}{c|}{}\\
$\eta$5e-2--$v$j3e3--n1e5--T1e3 & 50 & 3.2 & 1.4e-6 & 1e-7 &0.05 &1e4 &1e4 & 1e4 &6e3 &1.9e-6 &4.2e-3 &0.85 &2.4e-4\\
$\eta$5e-2--$v$j3e3--n1e5--T1e5 & 12 & 0.08 & 1e-8 & 8e-10 &0.05 &1e4 &1e4 & 1e4 &1e5 &4.4e-9 &2.9e-2 &2.0e-3 &5.5e-7\\
\hline 
\hline
\end{tabular}
}
\end{center}
\begin{flushleft}
This is a partial list of simulations studied here with different jet and background gas parameters. The columns list: 
(1) Model name:  The naming of each model starts with the feedback mass fraction, followed by the jet velocity in km s$^{-1}$ for kinetic jet or jet temperature in K for thermal dominant jet. The final 2 numbers labels the background gas density in cm$^{-3}$ and background gas temperature in K. 
(2) $\Delta T$: Simulation duration (all shorter than the free-fall time for constant $n_\infty$ gas without a BH). 
(3) Box size of the simulation.
(4) $m^{\rm max}_{\rm g}$: The highest mass resolution.
(5) $m^{\rm max}_{\rm jet}$: The mass resolution of the spawned jet particles.
(6) $\eta_{\rm m, fb}$: The feedback mass fraction.
(7) $V_{\rm jet}$: The initial jet velocity at spawn.
(8) $T_{\rm jet}$: The initial jet temperature at spawn.
(9) $n_\infty$: The background gas density.
(10) $T_\infty$: The background  gas temperature.
\ksu{(11) $\langle \dot{M}_{\rm BH} \rangle$: The resulting time-averaged accretion rate.
(12) $\langle \dot{M}_{\rm BH} \rangle/\dot{M}_{\rm Bondi}$: The same value over Bondi accretion rate.  
(13) $\langle \dot{M}_{\rm BH} \rangle/\dot{M}_{\rm Edd}$: The same value over the Eddington accretion rate ($\dot{M}_{\rm Edd}\equiv\dot{L}_{\rm Edd}/0.1 c^2$).     
(14) $\langle \dot{E}_{\rm jet} \rangle/\dot{E}_{\rm Edd}$: Jet energy flux over Eddington luminosity.}\\
$^*$ We run three variations of the same run with different random seeds \zh{for the stochastic injection of jet particles}
(labeled as I, II, III) to characterize \zh{the impact of this stochasticity}. Unless specified otherwise, in the rest of this paper we refer to run I.
\end{flushleft}
\end{table*}

\section{Two modes of jet propagation} \label{S:j_vel_difference}

Before exploring all of the simulations that we have run, we first focus on a set of three simulations all with the same fiducial background gas properties ($n_\infty=10^4 {\rm cm}^{-3}$ and $T_\infty=10^4$~K) and 
\zh{feedback mass fraction}
($\eta_{\rm m, fb}=0.05$), but varying the jet velocity, ranging from 3000 - 30000 km s$^{-1}$. These models are denoted as ``$\eta$5e-2--$v$j3e3--n1e5--T1e4'',  ``$\eta$5e-2--$v$j1e4--n1e5--T1e4'' (I,II,III), and ``$\eta$5e-1--$v$j3e4--n1e5--T1e4''. These \zh{velocity} variations result in very different jet cocoons and, as we will see, guide our development of a simple model which will explain how the jet evolves when we modify other parameters (such as the background gas properties).

\subsection{Cocoon morphology} \label{S:cocoon_v}
\begin{figure*}
    \centering
    \includegraphics[width=17cm]{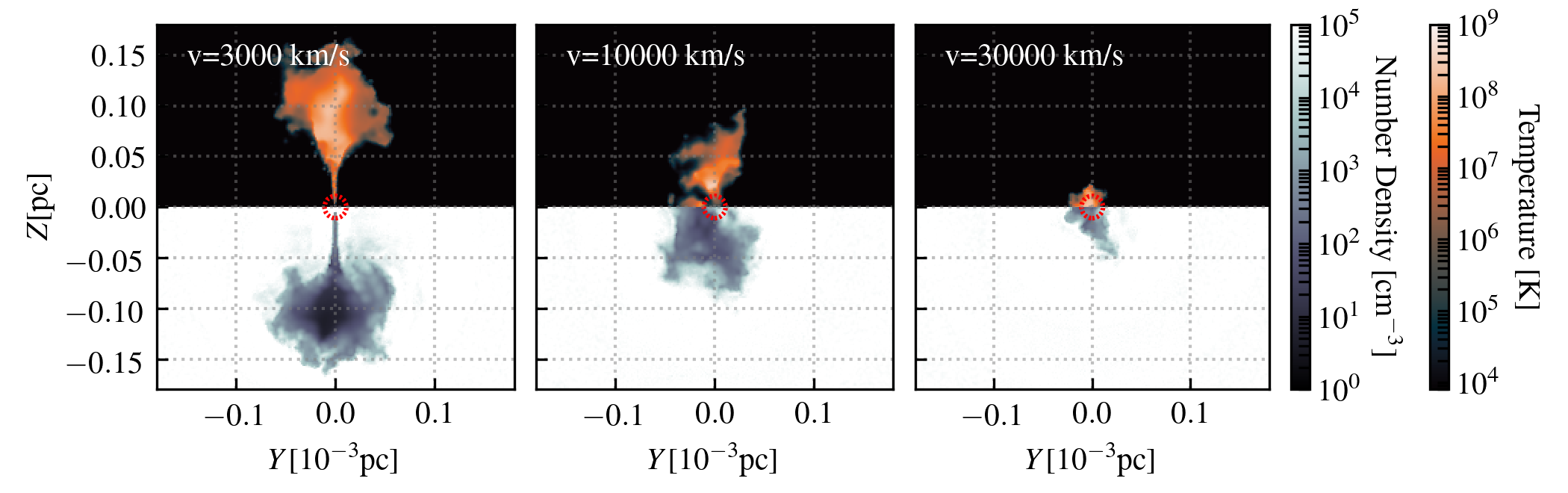}
    \caption{The number density (lower half of each panel) and temperature (upper half of each panel) morphology of the runs varying the jet velocity from 3000 to 30000 km s$^{-1}$($\eta$5e-2--$v$j3e3--n1e5--T1e4, $\eta$5e-2--$v$j1e4--n1e5--T1e4, and $\eta$5e-2--$v$j3e3--n1e5--T1e4 ) and keeping everything else constant. We 
    \zh{show 2D slices from 3D snapshots, selected at the time} when the cocoon has reached its maximum height. \ksu{The red dotted circle indicates the Bondi radius} in each panel. We see that low jet velocities result in elongated cocoons while high velocities produce a more spherical bubble-like morphology, and propagate to a shorter distance.  }
    \label{fig:cocoon_v}
\end{figure*}

\fref{fig:cocoon_v} shows the morphology of the cocoon for the three different jet velocities, depicting the resulting density and temperature distributions. 
Note that, in these figures, the hot jet gas is most clearly visible, but this is surrounded by a region of shocked ambient material. We use the term ``cocoon'' to refer to the combination of both regions. 
\ksu{The black hole accretion and resulting jet are highly episodic (see \fref{fig:flux_v}), and as a result, the length and the width of the jet cocoon are also time-dependent.}  We choose a snapshot where each cocoon reaches its maximum height in order to show the differences most clearly. This figure shows that the propagation of the jet cocoon varies primarily in length. The run with a lower jet velocity has a much more elongated jet cocoon, reaching a much larger distance. On the other hand, the higher velocity runs result in a roughly isotropic bubble-shaped cocoon. The higher the velocity, the shorter the distance the jet cocoon reaches.

Qualitatively, this is primarily because a lower velocity jet with lower specific energy regulates itself to a higher mass and momentum flux (for reasons we will discuss below). The 
higher mass and momentum flux jet can punch through to a much larger radius, 
consistent with what we see in galaxy scale jet simulations (e.g., \citealt{2003A&A...398..113K}, \citealt{2015ApJ...803...48G}, \citealt{2021MNRAS.507..175S} and Weinberger et al. in prep.). We will provide a more quantitative scaling for the propagation of the jet cocoon \zh{with jet models and the initial external gas density and temperature} in \sref{S:toy}. 

\subsection{BH Accretion Rate and Jet energy Flux} \label{S:flux_v}

\begin{figure}
    \centering
    \includegraphics[width=8cm]{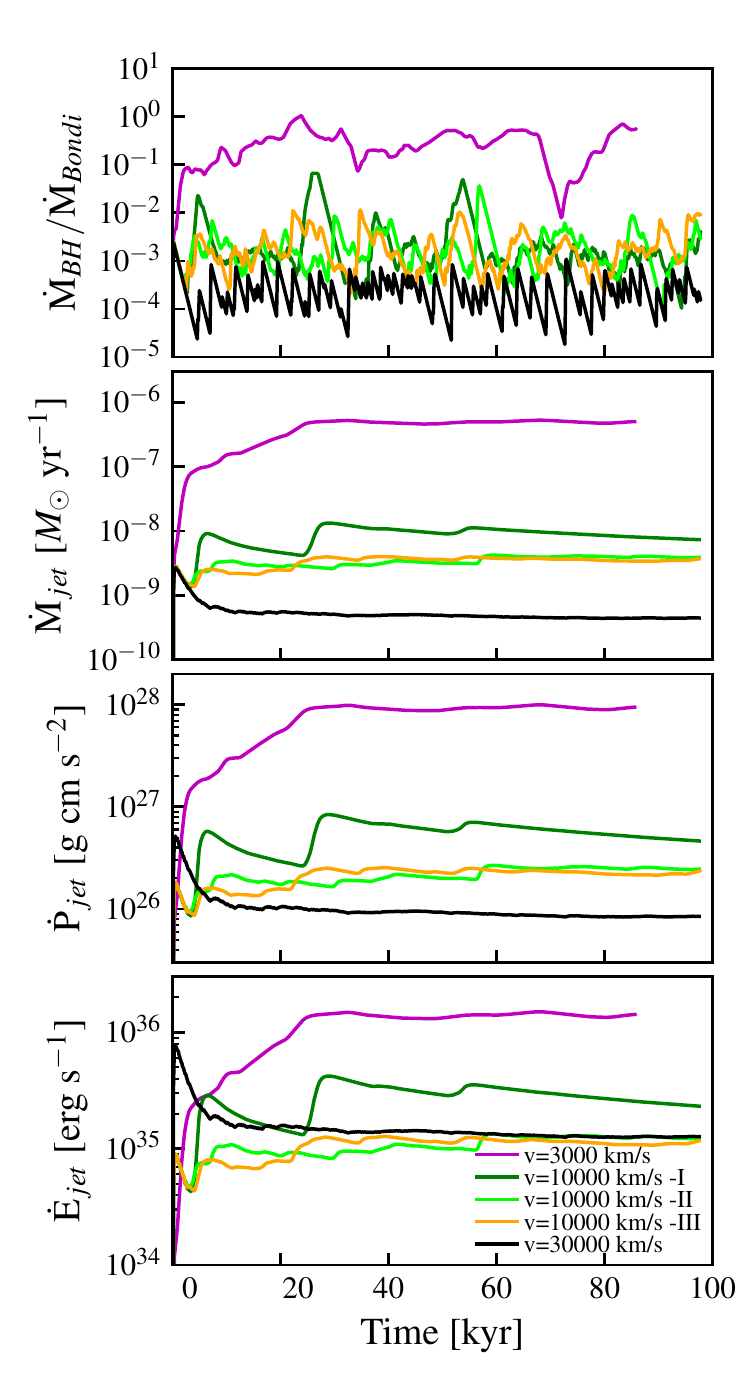}
    \caption{\zh{From top to bottom, the panels show (i) the black hole accretion rate, (ii) jet mass flux, (iii) momentum flux and (iv) energy flux in} runs varying the jet velocity from 3000 to 30000 km s$^{-1}$ (\zh{labeled as} ``$\eta$5e-2--$v$j3e3--n1e5--T1e4'', ``$\eta$5e-2--$v$j1e4--n1e5--T1e4'', and ``$\eta$5e-2--$v$j3e3--n1e5--T1e4'' \zh{in Table~\ref{tab:run}}). \ksu{Note that I, II, and III in the name label the same runs with different random seeds to show the range of stochastic variations. The 2nd, 3rd, and 4th panels show cumulative averages from the beginning of the run 
    up to the specific time of the run.} The lower velocity jet model results in a much higher black hole accretion rate and jet mass, momentum and energy fluxes. The higher velocity jets (10000 and 30000 km s$^{-1}$) regulate themselves to a similar jet energy flux values.  }
    \label{fig:flux_v}
\end{figure}

\fref{fig:flux_v} shows the resulting black hole accretion rate, jet mass flux, momentum flux and energy flux as a function of time for the same set of runs. The latter quantity is the cumulative-average from the beginning of the run $t=0$ to the given time to reduce the noise. 
With a feedback mass fraction of $\eta_{\rm m, fb}=0.05$, the black hole accretion rate roughly regulates to $2 \dot{M}_{\rm bond}$, $\sim 0.02-0.03 \dot{M}_{\rm bond}$, and  $\sim0.002 \dot{M}_{\rm bond}$ for jet velocities of 3000, 10000, and 30000 km/s, respectively.   We also see that the higher the jet velocity, the more short-term variability there is throughout the simulations, a topic we will return to in \sref{s:duty}.

Consistent with what we saw in \sref{S:cocoon_v}, a low-velocity jet regulates to a much higher mass and momentum flux, which is responsible for the more elongated cocoon. The 10000 and 30000 km s$^{-1}$ runs, both of which have cocoons that isotropize at small radius, roughly regulate to a similar jet energy flux, meaning that $\dot{M}_{\rm jet}\propto V_{\rm jet}^{-\kappa_v}$ with $\kappa_v\sim 2-2.5$. The lowest velocity run (3000 km s$^{-1}$), on the other hand, results in an even higher energy flux, qualitatively consistent with the much larger volume of the cocoon we see in \fref{fig:cocoon_v}. The lower velocity runs ($V_{\rm jet}\leq10000$ km s$^{-1}$) roughly have $\dot{M}_{\rm jet}\propto V_{\rm jet}^{-\kappa_v}$, with $\kappa_v\sim 3.5-4$. We will explore the reason behind the different behaviour and scalings of the high and low velocity jets in the next section.

\section{A simple model for jet propagation and cocoon formation } \label{S:toy}

In the previous section, we found that when we varied the jet velocity, the jets all self-regulated, but this could \zh{result either in a} nearly spherical, or in a highly elongated cocoon. Here we develop a simple analytic model based on this dichotomy and then, in the next section, we will use it to understand self-regulation when other parameters, such as the background gas properties, are changed.

\subsection{Jet propagation} \label{S:propogate}

\begin{figure*}
    \centering
    \includegraphics[width=18cm]{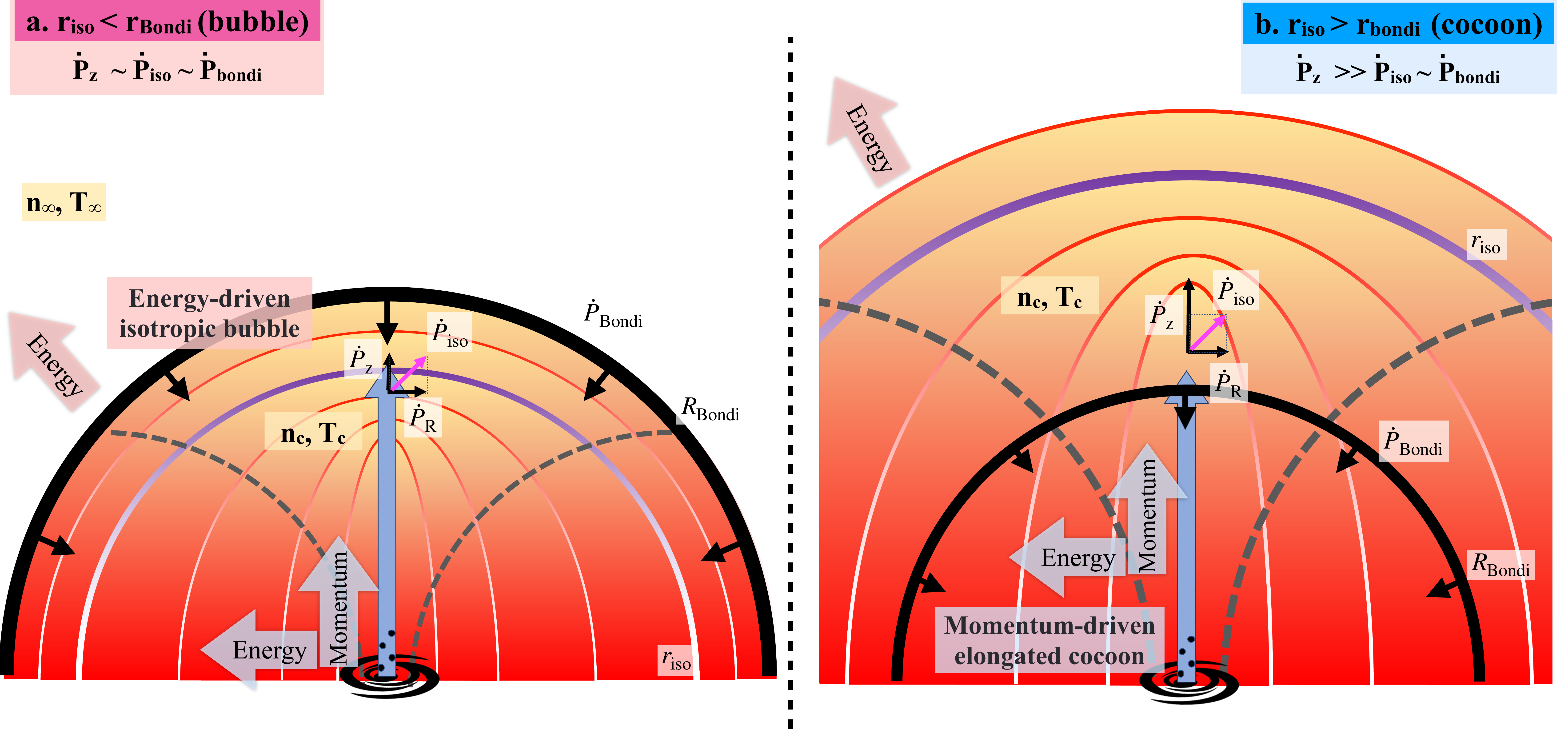}
    \caption{\ksuu{A cartoon picture of the jet cocoon propagation and the two possible cocoon morphologies as a result of different jet parameters, background gas density, and temperature. The left panel shows the isotropic ``bubble'' case where $r_{\rm iso}<r_{\rm Bondi}$. The right panel shows the elongated ``cocoon'' case where $r_{\rm iso}>r_{\rm Bondi}$. The blue arrow represents the jet. Each half oval indicates the jet cocoon at a given time. The gray dashed line indicates the resulting overall effective cocoon shape of a continuous jet injection by ``linking'' the cocoon shock-front at each time. }}
    \label{fig:cartoon}
\end{figure*}

We begin by reviewing the scaling which controls the cocoon shape before, in the next section, connecting this back to the accretion and hence overall self-regulation.

%
%
%
%
\subsubsection{Elongated jet cocoon -- before the cocoon isotropizes}
\ksu{We start by assuming the jet cocoon roughly follows a cylindrical geometry. As shown in \fref{fig:cartoon}, closer to where jets are launched,} the propagation of the jet qualitatively follows from momentum conservation in the z-direction \citep[e.g.,][]{1989ApJ...345L..21B,2021MNRAS.507..175S},
\ksuu{\begin{align}
\label{eq:mom_conserve}
 A_c V_z \rho_\infty V_z= 
 \frac{1}{2}\dot{M}_{\rm jet} V_{\rm jet},
 \end{align}}
 where $A_c = \pi R_{\rm cocoon}^2$ is the cross section of the pressurized cylinder (cocoon),  $V_z\equiv dz_{\rm cocoon}/dt$ is the expansion velocity of the cocoon in the polar directions,  $\dot{M}_{\rm jet}$ is the jet's initial mass flux, and $V_{\rm jet}$ is the initial jet velocity.

The evolution in the perpendicular direction is, instead, dictated by energy conservation, as the build-up of an over-pressured cocoon drives lateral expansion. \ksuu{The resulting expansion then pushes the surrounding gas.} \zh{The equations describing the conservation of energy and momentum flux can be written as}
\ksuu{\begin{align}
\label{eq:e_conserve}
A_{\rm tot}V_{R, {\rm Hot}}\left(\frac{1}{2}\rho_c V_{R, {\rm Hot}}^2\right) =& 
\frac{\gamma}{2}(\dot{M}_{\rm jet}\,V_{\rm jet}^2),\notag\\
V_{R, {\rm Hot}}^2\rho_c =& V_R^2\rho_\infty
\end{align}},
where  $A_{\rm tot} = 4 \pi\beta R_{\rm cocoon} z_{\rm cocoon}$ is the lateral surface area of the same region, $z_{\rm cocoon}$ is the height to which the jet reaches, \ksu{$V_{R, {\rm Hot}}$ is the immediate post shock velocity of the hot cocoon gas,} $V_R\equiv dR_{\rm cocoon}/dt$ is the expansion velocity of the cocoon in the mid-plane direction, $\beta$ is an order-of-unity geometric factor for the surface area of the cocoon \ksu{with respect to an ideal cylindrical geometry}, and $\gamma\equiv \dot{E}_{\rm expansion}/\dot{E}_{\rm kin}\propto \dot{E}_{\rm jet}/\dot{E}_{\rm kin}\equiv f_{\rm kin}^{-1}$ is the ratio of the energy flux in the perpendicular direction 
(proportional to the total injected energy $\dot{E}_{\rm jet}$) to the injected kinetic energy flux. Energy conservation is appropriate for the (initial) lateral expansion despite the strong cooling that can occur at the interface between the hot and cold gas within the cocoon. The total amount of cooling at this interface is proportional to its area (i.e. $A_c$) and so is negligible compared with the jet energy flux at early times. 

In this expression, $\rho_c$ is the cocoon gas density, which we will assume depends on the jet velocity and the background gas properties as:
\begin{align}
\label{eq:index}
\rho_c\propto\rho_\infty^\zeta T_\infty^\xi V_{\rm jet}^\delta.
\end{align}
where $\zeta$, $\xi$, and $\delta$ are exponents that we will determine later.
\ksu{Assuming the cocoon is pressurized by strong shocks (where $\rho_{\rm post}\sim 2\rho_{\rm pre}$ and $v_{\rm post}\sim 0.5 v_{\rm pre}$),  $\gamma$ is roughly 
\begin{align}
\gamma_{\rm super-sonic} &\sim \frac{\dot{E}_{\rm post-shock}}{\dot{E}_{\rm pre-shock}} \frac{\dot{E}_{\rm pre-shock}}{\dot{E}_{\rm jet}}\frac{\dot{E}_{\rm jet}}{\dot{E}_{\rm kin}}\notag\\
&\sim\frac{\rho_{\rm post} v_{\rm post}^3}{\rho_{\rm pre} v_{\rm pre}^3}\times(1-f_{\rm loss})f_{\rm kin}^{-1}\notag\\
&\sim \frac{1}{16} \times(1-f_{\rm loss})f_{\rm kin}^{-1}\,\,\lesssim\frac{1}{16}f_{\rm kin}^{-1}.
\end{align}
Therefore, we assume $\gamma$ is a constant for the remainder of the paper.
}

\ksuu{From the equations above, we can solve for the time dependence of $V_R$ and $R_{\rm cocoon}$ as
\begin{align}
\label{eq:V_r_time}
V_R=\left(\frac{\gamma^2}{72\pi\beta^2}\right)^{1/6}\dot{M}_{\rm jet}^{1/6}\rho_c^{1/6}\rho_\infty^{-1/3}V_{\rm jet}^{1/2}t^{-1/3}\notag\\
R_{\rm cocoon}=\left(\frac{81\gamma^2}{512\pi\beta^2}\right)^{1/6}\dot{M}_{\rm jet}^{1/6}\rho_c^{1/6}\rho_\infty^{-1/3}V_{\rm jet}^{1/2}t^{2/3}
\end{align}
and the time dependence of $V_z$ and $z_{\rm cocoon}$ as
\begin{align}
\label{eq:V_z_time}
V_z=\left(\frac{8\beta}{9\pi\gamma}\right)^{1/3}\dot{M}_{\rm jet}^{1/3}\rho_c^{-1/6}\rho_\infty^{-1/6}t^{-2/3}\notag\\
z_{\rm cocoon}=\left(\frac{24\beta}{\pi\gamma}\right)^{1/3}\dot{M}_{\rm jet}^{1/3}\rho_c^{-1/6}\rho_\infty^{-1/6}t^{1/3}.
\end{align}
}

In particular, the opening angle of the resulting cocoon scales as 
\begin{align}
\label{eq:angle_z}
\frac{R_{\rm cocoon}}{z_{\rm cocoon}}&= \frac{\gamma z_{\rm cocoon}}{16\beta}\left(\frac{2\pi \rho_c V_{\rm jet}}{\dot{M}_{\rm jet}}\right)^{1/2}.
\end{align}
We have assumed that the jet starts such that $R_{\rm cocoon} < z_{\rm cocoon}$, but as the cocoon propagates, for a fixed $\dot{M}_{\rm jet}$, eventually, it becomes (quasi-)isotropic ($R_{\rm coocoon}\sim z_{\rm cocoon}$); this occurs at a time given by
\ksuu{\begin{align}\label{eq:t_iso}
t_{\rm iso}=&\left(\frac{2^{15}\dot{M}_{\rm jet}\rho_\infty}{\pi V_{\rm jet}^3\rho_c^2}\right)^{1/2}\left(\frac{\beta^2}{3\gamma^2}\right)\notag\\
\approx &660 {\rm yr} \times (1-f_{\rm loss})^{-2} f_{\rm kin}^2 \left(\frac{\bar{n}}{10^5 {\rm cm}^{-3}}\right)^{-1/2}\notag\\
&\left(\frac{\dot{M}_{\rm jet}}{5\times10^{-9}\,{\rm M}_\odot {\rm yr}^{-1} }\right)^{1/2} 
\left(\frac{V_{\rm J}}{10^4\,{\rm km\, s}^{-1} }\right)^{-3/2},
\end{align}
}
and at a cocoon height of (see \fref{fig:cartoon})
\ksu{
\begin{align}\label{eq:z_iso}
&z_{\rm iso}\equiv r_{\rm iso}=\left(\frac{\dot{M}_{\rm jet}}{2\pi\rho_c V_{\rm jet}}\right)^{1/2}\left(\frac{16\beta}{\gamma}\right)\notag\\
&\approx 1.3\times10^{-3} {\rm pc}\,\,\, (1-f_{\rm loss})^{-1}f_{\rm kin}\times \left(\frac{\bar{n}}{10^{5}\,{\rm cm}^{-3} }\right)^{-1/2} \notag\\
&\left(\frac{\dot{M}_{\rm jet}}{5\times10^{-9}\,{\rm M}_\odot {\rm yr}^{-1} }\right)^{1/2}\left(\frac{V_{\rm J}}{10^4\,{\rm km\, s}^{-1} }\right)^{-1/2}.
\end{align}
}
\subsubsection{Isotropic bubble -- after the cocoon isotropizes}

After the cocoon isotropizes, the momentum no longer dominates the jet propagation as $V_R$ grows larger than $V_z$. The whole cocoon becomes an energy-driven expanding bubble as shown in the outer part of \fref{fig:cartoon}.
In this case,
\ksuu{
\begin{align}
\label{eq:e_conserve_iso}
 &4\pi R^2\rho_c V_{R, {\rm Hot}}^3=\frac{\gamma'}{2} \dot{M}_{\rm jet} V_{\rm jet}^2\notag\\
&V_{R, {\rm Hot}}^2\rho_c=V_R^2\rho_\infty,
\end{align}
}
where $\gamma'\equiv \dot{E}_{\rm expansion}/\dot{E}_{\rm kin}$.
\ksuu{Note that this matches \Eqref{eq:e_conserve} assuming $R_{\rm cocoon}=z_{\rm cocoon}$ up to an order-of-unity geometry factor, which we treat in a very approximate manner. Again assuming supersonic shocks, then $\gamma'\sim\gamma_{\rm super-sonic}$. \Eqref{eq:e_conserve_iso} has the solution:
\begin{align}\label{eq:iso_r_time}
V_{\rm R}=\left(\frac{9\gamma'\dot{M}_{\rm jet}V_{\rm jet}^2\rho_c^{1/2}}{200\pi\rho_\infty^{3/2}}\right)^{1/5}
 (t-t_{\rm iso})^{-2/5}\notag\\
R=\left(\frac{125\gamma'\dot{M}_{\rm jet}V_{\rm jet}^2\rho_c^{1/2}}{216\pi\rho_\infty^{3/2}}\right)^{1/5}
 (t-t_{\rm iso})^{3/5}.
 \end{align}
}

\subsection{Feedback self-regulation} \label{S:self_reg}

Turning to the physics of self-regulation, we note that, at the Bondi radius $R_{\rm Bondi}=GM_{\rm BH}/c_s^2$, the inflowing mass flux goes as 
\begin{align}
\dot{M}_{\rm Bondi}= \frac{e^{3/2}\pi \rho_\infty G^2 M^2}{c_s^3}
\end{align}
and the inflowing momentum flux goes as
\begin{align}
\label{eq:mom_inflow}
\dot{P}_{\rm Bondi}=\dot{M}_{\rm Bondi} V_{\rm ff}|_{R_{\rm Bondi}}= \frac{e^{3/2}\pi \rho_\infty G^2 M_{\rm BH}^2}{c_s^2}.
\end{align}

Regulation will occur when the jet cocoon produces a momentum flux which matches this.
\ksu{However, if the momentum flux is very anisotropic such that the $z$ component of momentum flux $\dot{P}_{\rm z,cocoon}$ is much larger than the momentum flux perpendicular to the jet $\dot{P}_{\rm R,cocoon}$ , the extra momentum in the z-direction is insufficient, by itself, to stop the accretion.}
Therefore, we argue that regulation happens when the isotropic component of the jet cocoon or bubble momentum flux matches the inflowing momentum flux at the Bondi radius.
\begin{align}
\label{eq:match}
    &4\pi \rho_\infty R_{\rm Bondi}^2  V_{\rm iso,Bondi}^2 = \frac{e^{3/2}\pi \rho_\infty G^2 M_{\rm BH}^2}{c_s^2},
\end{align}
where $V_{\rm iso,Bondi}$ is the isotropic component of the cocoon velocity at the Bondi radius. \ksu{We estimate the isotropic component of the cocoon velocity as $V_{\rm iso}\equiv \sqrt{2}\min(V_R,V_Z)$}. We explain how we estimate its value under different conditions as follows.

\subsubsection{$z_{\rm iso}>R_{\rm Bondi}$}

As shown in the right part of \fref{fig:cartoon}, if the jet cocoon isotropizes at a radius larger than the Bondi radius ($z_{\rm iso}>R_{\rm Bondi}$), $V_z>V_R$ when $R$ reaches $R_{\rm Bondi}$, we estimate the isotropic component of velocity at the Bondi radius to be 
\begin{align}\label{eq:V_c_iso}
V_{\rm iso,Bondi}&=\sqrt{2}V_R|_{R_{\rm Bondi}}\notag\\
&\propto \dot{M}_{\rm jet}^{1/4}\,V_{\rm jet}^{3/4}\,R^{-1/2}_{\rm Bondi}\,\rho_c^{1/4} \rho_\infty^{-1/2}.
\end{align}
From \Eqref{eq:match} and \Eqref{eq:V_c_iso}, we find that the jet mass flux regulates to
\begin{align}\label{eq:reg_cocoon}
\dot{M}_{\rm jet} &\propto M_{\rm BH}^2\rho_\infty^2\rho_c^{-1}V_{\rm jet}^{-3}\notag\\
& \propto  M_{\rm BH}^2 \rho_\infty^{2-\zeta} \,\,T_\infty^{-\xi}\,\, V_{\rm jet}^{-3-\delta}.
\end{align}

\subsubsection{$z_{\rm iso}<R_{\rm Bondi}$}

On the other hand, as shown in the left part of \fref{fig:cartoon}, if $z_{\rm iso}<R_{\rm Bondi}$, the cocoon can also become an isotropic bubble before reaching the Bondi radius.
Therefore, \Eqref{eq:e_conserve_iso} in this case gives
\ksuu{
\begin{align}\label{eq:V_b_iso}
V_{\rm iso,Bondi}= \left(\frac{\gamma' \dot{M}_{\rm jet} V_{\rm jet}^2\rho_c^{1/2}}{8\pi \rho_\infty^{3/2} R_{\rm Bondi}^2}\right)^{1/3}.
\end{align}}
From \Eqref{eq:match} and \Eqref{eq:V_b_iso}, we see that the jet mass flux in this case regulates to
\begin{align}\label{eq:reg_bubble}
\dot{M}_{\rm jet}&\propto M_{\rm BH}^2\rho_\infty^{3/2}\rho_c^{-1/2}V_{\rm jet}^{-2} c_s^{-1}\notag\\
&\propto M_{\rm BH}^2\rho_\infty^{(3-\zeta)/2}V_{\rm jet}^{-2-\delta/2} T_\infty^{-(1+\xi)/2}.
\end{align}

\subsection{Cocoon or Bubble at the Bondi Radius?}
The jet cocoon will be elongated at the Bondi radius if $z_{\rm iso}>R_{\rm Bondi}$ or, from \eeqref{eq:z_iso}, if 
\begin{align}\label{eq:criteria}
\dot{M}_{\rm jet} > \left(\frac{\pi \gamma^2 R_{\rm Bondi}^2}{128 \beta^2}\right) \rho_c V_{\rm jet}\sim  \rho_\infty^\zeta\,\, T_\infty^\xi V_{\rm jet}^{1+\delta}.
\end{align}
Otherwise, it isotropizes before reaching the Bondi radius.\ksu{ We next list which of these two scenarios is realized for different parameter values, as follows.}

\begin{itemize}
   \item {\bf Jet velocity:\,\,\,}\label{s:gas_c}
This mass flux criterion scales as $V_{\rm jet}$, but the mass fluxes are regulated to a negative power of $V_{\rm jet}$ in both the cocoon (\eeqref{eq:reg_cocoon}) and bubble (\eeqref{eq:reg_bubble}) cases, as $\delta$ is small (as we measured in simulation). 
Therefore the lower the jet velocity, the narrower the cocoon at the Bondi radius.

\item {\bf Gas density:\,\,\,}\label{s:gas_n}
If $\zeta$ (from \eeqref{eq:index}) is smaller than 1 (which is the case, as will be shown later in \sref{s:cocoon_phase}), then $\dot{M}_{\rm jet}$ has a super-linear dependence on $n_\infty$ for both the elongated cocoon and isotropic bubble cases. The separation between the two cases, on the other hand, has $\dot{M}_{\rm jet}$ scaling linearly with $n_\infty$ (\eeqref{eq:criteria}). From the same argument as above, the higher the background density, the more elongated the jet cocoon is.

\item{\bf Gas temperature:\,\,\,}\label{s:gas_t}
If $\xi$ (from \eeqref{eq:index}) is close to zero (which is the case as will be shown later in \sref{s:cocoon_phase}), the $\dot{M}_{\rm jet}$ in the elongated cocoon case will have little dependence on $T_\infty$, while the bubble case will have a scaling of $\dot{M}_{\rm jet}\propto T_\infty^{-0.5}$. 
The separation between the two cases has a negligible dependence of $\dot{M}_{\rm jet}$ on $T_\infty$, (\eeqref{eq:criteria}). From the same argument above, if the background temperature increases, the cocoon shape either remains the same or becomes slightly more bubble-like.
\end{itemize}


\section{Simulation results: Cocoon regulation and black hole accretion} \label{s:result}

Armed with a better understanding of the physics of jet regulation from the simple scalings obtained in the previous section, we next turn to a more complete examination of the simulation results. We begin by demonstrating that the isotropic momentum is key to self-regulation, before discussing first the cocoon's properties, and then the black hole accretion rate and growth.

\subsection{The \zh{self-regulation of the cocoon by its isotropic momentum flux} }
\begin{figure*}
    \centering
    \includegraphics[width=17cm]{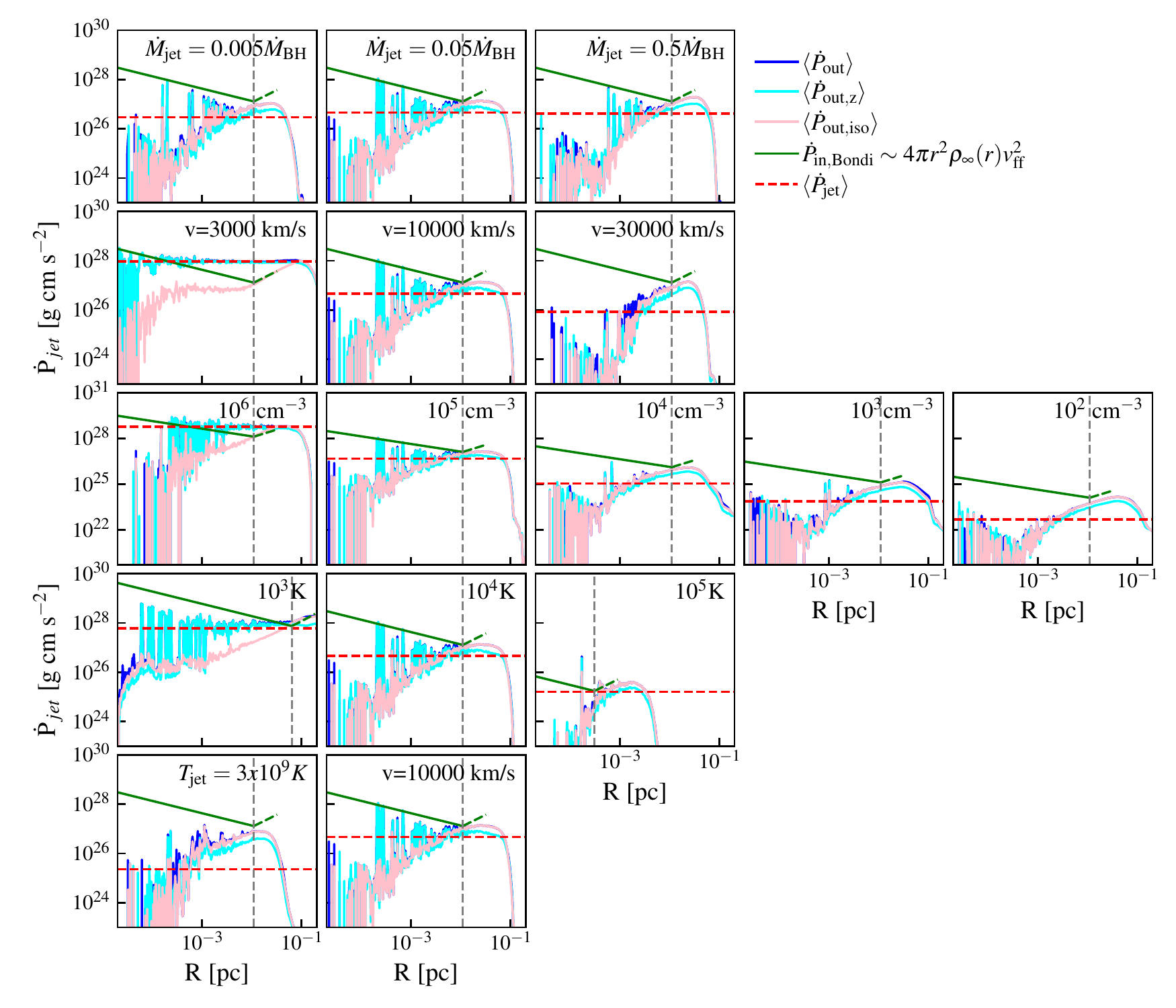}
    \caption{The comparison of time-averaged momentum fluxes from the simulations. Three types of momentum flux are shown: (i) the average jet momentum flux (red), (ii) the cocoon momentum flux, with blue, cyan, and pink lines showing the total cocoon flux, its $z$ component, and the isotropic 
    component \ksu{($\min(\dot{P}_{\rm out}, 2\dot{P}_{\rm out, R},2\dot{P}_{\rm out, z})$, see \Eqref{eq:cocoon_mom} and \fref{fig:cartoon})}, respectively, and (iii) finally the estimated inward Bondi momentum flux (green). The vertical line in each plot marks the Bondi radius. The isotropic component of the outward cocoon momentum flux matches the inward Bondi momentum flux at the Bondi radius. Runs with elongated cocoons (v=3000 km s$^{-1}$ and $n=10^6$ cm$^{-3}$) have the $z$-component of their cocoon fluxes roughly match the jet momentum fluxes (momentum-driven) and are much larger than the isotropic components. Runs with bubble-shaped cocoons all have their cocoon momentum fluxes (energy-driven) higher than the jet momentum fluxes. }
    \label{fig:reg}
\end{figure*}

\begin{figure*}
    \centering
    \includegraphics[width=18cm]{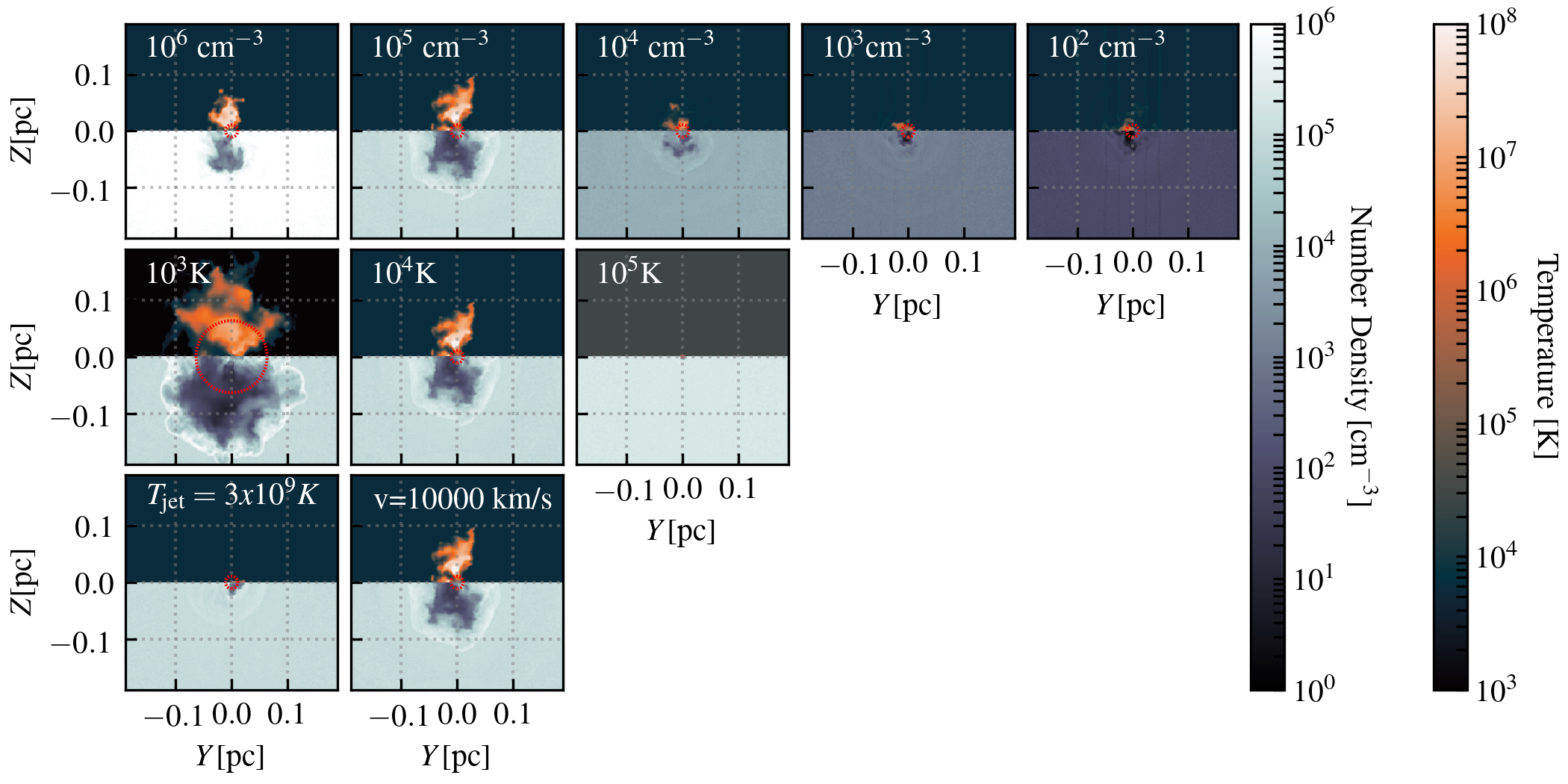}
    \caption{The density and temperature morphology of the other runs varying the other parameters. Higher-density runs have \zh{much larger and} more elongated jet cocoons. Lower-temperature runs have a much larger Bondi radius, so the cocoon also reaches further. A thermal energy dominant jet results in a \zh{quasi-spherical} bubble-shaped cocoon. }
    \label{fig:cocoon_all}
\end{figure*}

We first explicitly demonstrate that, in the simulations, the isotropic component of the cocoon momentum flux \ksu{(as defined in \eeqref{eq:V_c_iso} and \eeqref{eq:V_b_iso})} is roughly regulated to the inflow momentum flux, assuming a Bondi value. Each row in \fref{fig:reg} shows the variation of a specific parameter ($\dot{M}_{\rm jet}$, $v_{\rm jet}$, $n_\infty$, $T_\infty$, and $T_{\rm jet}$). There are three kinds of momentum flux plotted in each panel. The first is the injected jet momentum, time-averaged over the duration of each run, which is colored red. 

The second kind of momentum flux is the time-averaged cocoon momentum flux. The blue line shows its total value, the pink line the isotropic component,  and the cyan line the $z$ component. 
\ksuu{More specifically,  we define the cocoon momentum flux by summing the gas particles as
\begin{align}\label{eq:cocoon_mom}
&\langle P_{\rm out}    \rangle=\sum\limits_{r_i = r\pm\delta r, T_i>1.2 T_\infty, v_{r,}>0} \frac{m_i v_{r,i}^2}{\delta r}\notag\\
&\langle P_{\rm out, z} \rangle=\sum\limits_{r_i = r\pm\delta r, T_i>1.2 T_\infty, v_{z,i}\cdot sign(z)>0} \frac{m_i v_{z,i}^2}{\delta r}\notag\\
&\langle P_{\rm out, R} \rangle =\sum\limits_{r_i = r\pm\delta r, T_i>1.2 T_\infty, v_{r2d,i}>0} \frac{m_i v_{r2d,i}^2}{\delta r}\notag\\
&\langle P_{\rm out, iso}\rangle = \min(P_{\rm out}, 2P_{\rm out, R},2P_{\rm out, z}),
\end{align}
where $m_i$ is the particle mass, $T_i$ is the temperature, $r_i$ is the particle's 3D radial position, $v_{r,i}$ is the 3D radial velocity,  $v_{z,i}$ is the $z$-velocity, and $v_{r2d,i}$ is the lateral velocity. }

The third kind of momentum flux is the estimated inflowing momentum flux assuming a Bondi density profile 
\begin{align}
\rho \sim \begin{cases}
      \rho_\infty  & \text{for\,\,} r>R_{\rm Bondi}\\
      \rho_\infty \left(\frac{r}{R_{\rm Bond}}\right)^{-3/2}  & \text{for\,\,} r<R_{\rm Bondi}
    \end{cases} 
\end{align}
multiplied by the $4 \pi r^2 v_{\rm ff}^2$. \ksu{Note that $v_{\rm in}\sim V_{\rm ff}$ does not hold far beyond $R_{\rm Bondi}$, so we only plot this curve out to $\sim3 R_{\rm Bondi}$.} We immediately see that the isotropic component of the momentum flux (pink curves) is roughly regulated to the Bondi inflowing momentum flux (green curves) at the Bondi radius (vertical line). More specifically, the runs can be separated into two categories ---  $z_{\rm iso}>R_{\rm Bondi}$ (cocoon-like at $R_{\rm Bondi}$) and $z_{\rm iso}<R_{\rm Bondi}$ (bubble-like at $R_{\rm Bondi}$). 

Cross-referencing the morphological plots in  \fref{fig:cocoon_v} (for the simulations with jet velocity variation) and \fref{fig:cocoon_all} (for the simulations varying $n_\infty$, $T_\infty$, and $T_{\rm jet}$), both the $3000$ km s$^{-1}$ and $n=10^6\, {\rm cm}^{-3}$ runs fall clearly in the first category (elongated cocoons). In these runs, the $z$-direction momentum flux is roughly equivalent to the jet momentum flux, indicating a momentum-conserving propagation. Both are much larger than the isotropic component of the cocoon momentum flux until well beyond the Bondi radius, where the two components become comparable. The $z$-direction momentum flux is also larger than the inflowing momentum flux (assuming a Bondi value) at the Bondi radius. The isotropic component of the cocoon momentum flux, on the other hand, matches the inflowing momentum flux. In fact, they not only match at the Bondi radius, but they also match until the jet cocoon isotropizes, at a several times larger distance. 
This is primarily because the isotropic component of the velocity roughly scales as $r^{-1/2}$ (see \eeqref{eq:V_c_iso}), identical to the scaling of the free-fall velocity. 

The higher-velocity runs ($V_{\rm jet}>10000\, {\rm km\,  s}^{-1}$), lower-density runs ($n_\infty\lesssim 10^4 {\rm cm}^{-3}$) and thermal jet runs clearly fall in the second category (see also \fref{fig:cocoon_v} and \fref{fig:cocoon_all}). In this scenario, the cocoon isotropizes at a radius much smaller than the Bondi radius, and the isotropic component and the $z$ component become comparable over most of the plotted range. They are both larger than the input jet momentum flux as the propagation is energy-driven \zh{(i.e. by the thermal velocity, rather than the jet's bulk velocity; see \sref{S:propogate})}. However, they still match the inflowing momentum flux assuming the Bondi value.

The regulation of the isotropic component of the cocoon momentum flux to the Bondi value at $R_{\rm Bondi}$ is clearly reproduced in these results. When changing the background gas temperature by two orders of magnitude, the Bondi radius also differs by two orders of magnitude, and the two values still match.

\subsection{\zh{Thermal} phase structure of the cocoon/bubble gas}\label{s:cocoon_phase}

Before jumping into the implications of this regulation for black hole accretion, we will first need to understand how the cocoon phase structure depends on the jet model and gas properties. This is reflected in the power-law index in \Eqref{eq:index} and enters the regulation of the jet mass flux and accretion rate in \Eqref{eq:reg_cocoon} and \Eqref{eq:reg_bubble}. 

\begin{figure}
    \centering
    \includegraphics[width=8cm]{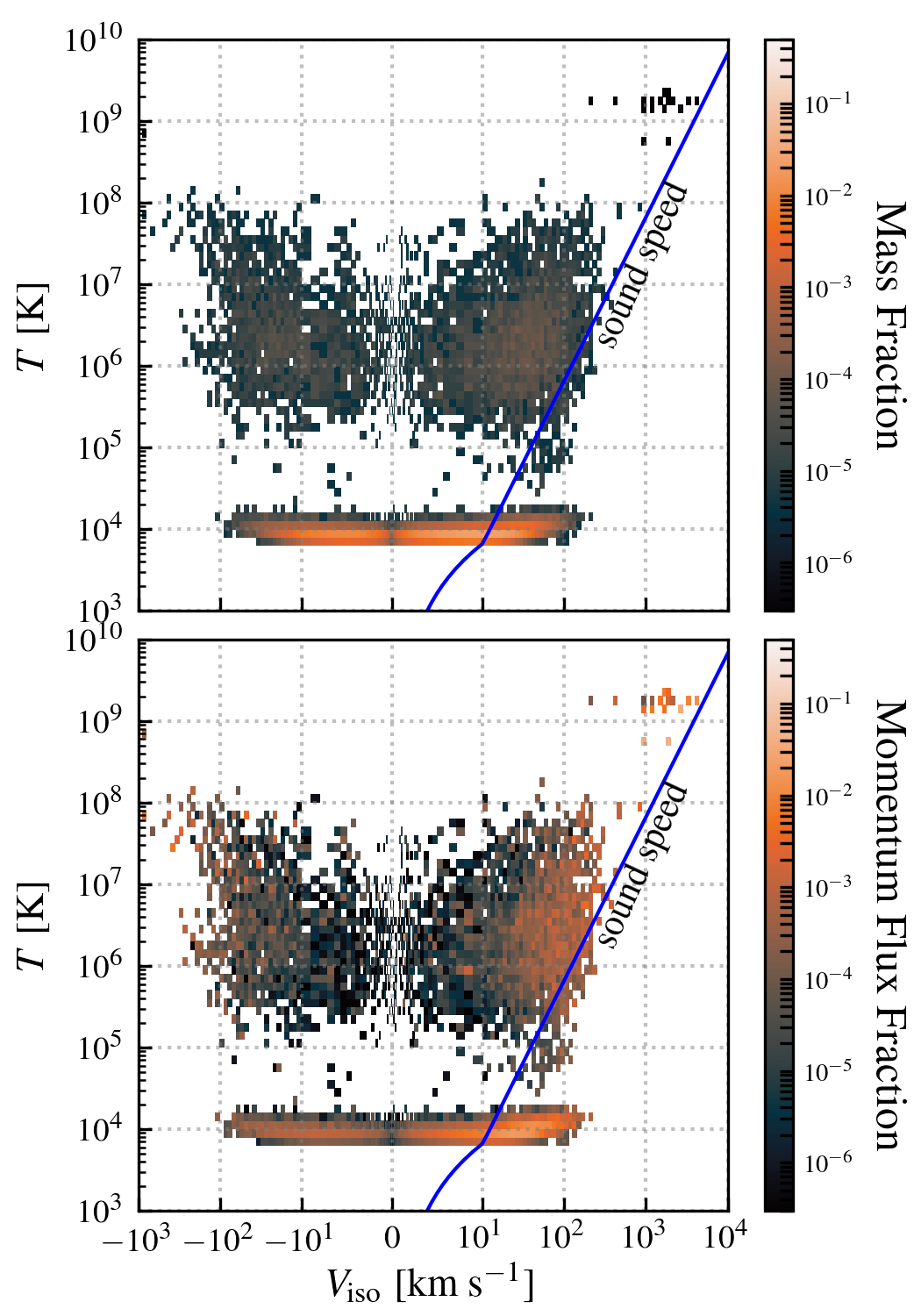}
    \caption{The phase structure of the cocoon gas ($T>1.2\,T_\infty$) in the temperature {\it vs.} isotropic velocity component plane ($V_{\rm iso}\sim \sqrt{2}V_{\rm 2d}$). The top panel is mass-weighted, and the lower panel is momentum-flux-weighted (isotropic component). The cocoon consists of a reverse-shocked, trans- to subsonic, turbulent hot phase and a cold mixing layer phase. The two phases contribute roughly equally to the isotropic cocoon momentum flux. Most of the mass, on the other hand, is in the cold phase. The hottest temperatures roughly correspond to the shock temperature implied by the jet velocity.}
    \label{fig:phase}
\end{figure}

\fref{fig:phase} shows the phase structure of the fiducial run $\eta$5e-2--$v$j1e4--n1e5--T1e4 in the temperature -- $V_{\rm iso}$ (isotropic component of cocoon velocity) plane. The top panel is mass-weighted, showing the phase distribution, while the bottom panel is momentum-flux-weighted, showing which phase contributes the most to the outflowing momentum flux. There are clearly two phases present. The first is the hot phase, which consists of the reverse-shocked hot gas filling the volume of the cocoon, and is primarily trans- to subsonic-turbulent. The second, colder, phase is roughly at the background gas temperature and density. The gas in this phase is at the ``mixing layer'' of the cocoon and surrounding gas, which is already cold. The second panel shows that both phases have a roughly equivalent contribution to the outflowing cocoon momentum flux, while most of the mass is in the cold phase.

\begin{figure}
    \centering
    \includegraphics[width=8cm]{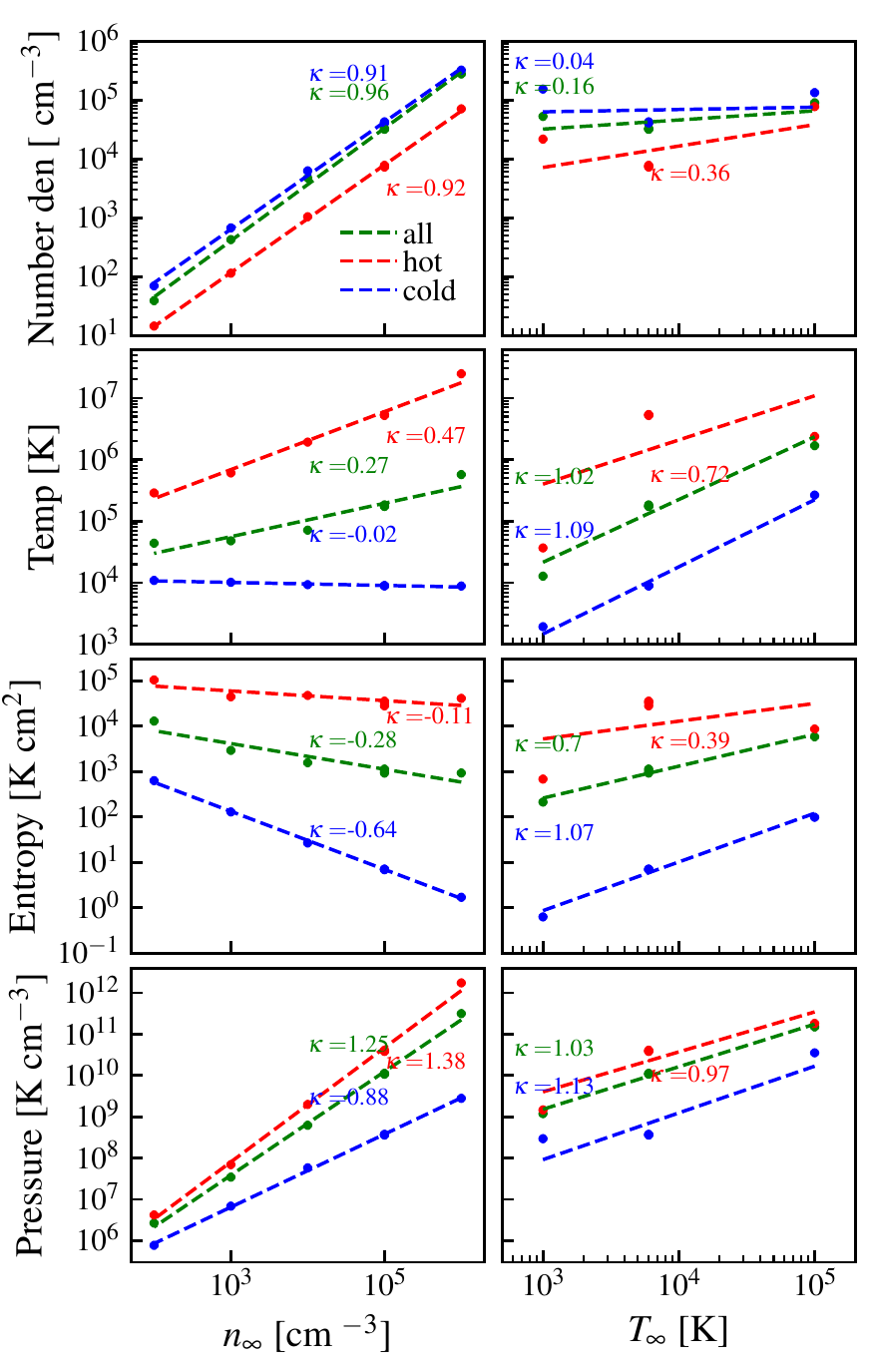}
    \caption{The dependence of the cocoon gas density, temperature, entropy, and pressure on background gas properties \zh{(the latter evaluated at the Bondi radius)}. The red, blue, and green dots and lines correspond to the hot, cold, and combined phases. The dots are from each simulation and the lines are fitted power laws with the index ($\kappa$) labeled. The cocoon is defined \zh{for simplicity} as all gas with $T > 1.2 T_\infty$.  We find $n_c\propto n_\infty^\zeta T_\infty^\xi$ with $\zeta\lesssim 0.9$ and $\xi\sim 0$.} 
    \label{fig:phase_fit}
\end{figure}

\fref{fig:phase_fit} shows an estimate of how the cocoon gas properties depend on the background gas properties \ksu{at the Bondi radius}. This is represented as $\zeta$ and $\xi$ in \Eqref{eq:index}. We note that the dependence on the jet velocity is weak ($\delta\sim0$), so we do not explicitly show it here. Given what we saw in \fref{fig:phase}, we fit for the gas properties of the whole cocoon (estimated as $T>1.2\, T_\infty$, shown with green lines), the cool-mixing-layer phase \ksu{($1.2\,T_\infty<T<3.6\,T_\infty$; blue lines)}, and the hot cocoon gas \ksu{($T>3.6\,T_\infty$; red lines)}. We include only gas with $V_{\rm iso}>0$. While averaging the cocoon gas properties, we volume-weighted the density and pressure while mass-weighting the temperature and entropy. We emphasize that this yields only an approximate estimate of the power law index, as each jet model goes through multiple cycles of feedback and the cocoon consists of multi-phase gas. To look at overall behaviour, we average over all times and the multi-phase cocoon gas at the Bondi radius, and fit a straight line through the results (in logarithmic quantities). 

We see from the left panel that the cold-mixing-layer gas generally follows the background gas temperature and density. On the other hand, the hot cocoon gas follows a constant entropy trend, as the reverse-shock heated gas has its properties set largely by the jet model instead of the background gas properties. We find a scaling approximately $n_c\propto n_\infty^\zeta$ with $\zeta\lesssim0.9$, consistent with our claim in \sref{s:gas_n} that the higher the background density, the more elongated the cocoon.

The right panel shows that the cocoon gas, in either phase, scales only weakly with the background temperature. Again, the cold-mixing-layer gas roughly matches the background gas temperature since they have both already cooled to the temperature floor ($T_\infty$). On the other hand, the hot phase has a steeper than linear scaling with the background temperature. Overall, we find a scaling of $n_c\propto T_\infty^\xi$ with $\xi\lesssim 0$. \ksu{This is also roughly consistent with our claim in \sref{s:gas_t} that, if $\xi$ is smaller than 0, the higher the background temperature, the more ``isotropic'' the jet cocoon (\sref{s:gas_t}).}

\subsection{The black hole accretion rate and jet mass flux}

\begin{figure}
    \centering
    \includegraphics[width=8cm]{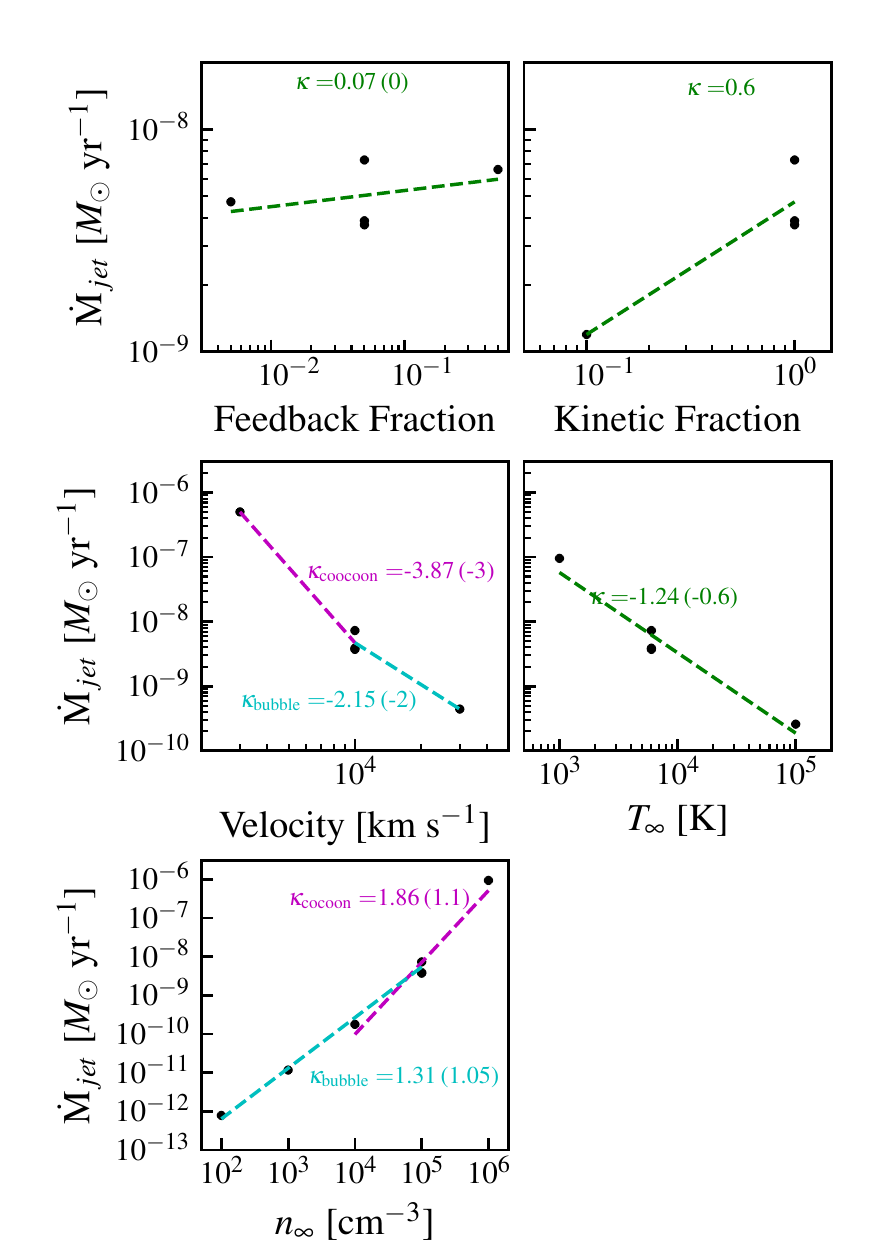}
    \caption{The dependence of the jet mass flux ($\dot{M}_{\rm jet}$) on the adopted jet model and background gas properties. The lines show power-law fits, with the index ($\kappa$) labeled. The number in the parenthesis is an estimate from the toy model and the fit to the cocoon gas-phase dependence in \fref{fig:phase_fit}. 
    }
    \label{fig:mdot_reg}
\end{figure}

Having determined how the cocoon properties depend on the background gas density and temperature, we can finally see whether the implied regulation of our jet and the resulting black hole accretion rate in our simple model can qualitatively explain what we see in the simulations.  \fref{fig:mdot_reg} shows the time-averaged jet mass flux in all of the runs, with each panel showing the variation of a specific parameter.    
We plot only the jet mass flux for ease of comparison with our simple model, but it should be kept in mind that this is directly proportional to the black hole accretion rate (simply scaled by a constant factor $\eta_{\rm m,fb}^{-1}$, which is $\sim 20$ for most runs).

The first panel shows that the jet mass flux is independent of the feedback mass fraction. The jets in these runs have the same specific energy, so the same jet mass flux means the same momentum and energy flux, which implies a very similar cocoon propagation.
With the lower feedback mass fraction, the black hole accretes more to provide an equivalent level of feedback. This holds until the required accretion rate is much larger than the Bondi accretion rate, in which case the jet model will fail to self-regulate. That scenario is not within the parameter space we simulate here.

In the second panel, we vary the kinetic fraction by varying the jet temperature and velocity while keeping the total specific energy the same. The lower kinetic fraction run has most of the energy in a thermal component, isotropizing the cocoon essentially at the launch of the jet. Moreover, since its cocoon has never been in a momentum conserving phase, it does not reach far beyond the Bondi radius. It can clearly be seen in the last rows of \fref{fig:reg} and \fref{fig:cocoon_all} that, although the isotropic component of the cocoon momentum flux matches the Bondi value at the Bondi radius in both the thermal and kinetic jets, the cocoon momentum flux decays more steeply beyond the Bondi radius. As a result, much less energy is ``wasted'' beyond the Bondi radius, so both the jet mass flux and black hole accretion rates regulate to a lower value.

In the simulations where we vary the jet velocity (center-left panel), the cocoon density depends weakly on the jet velocity, as mentioned in \sref{s:cocoon_phase}. Therefore \Eqref{eq:reg_cocoon} ($z_{\rm iso}>R_{\rm Bondi}$) predicts a scaling of $\dot{M}\propto V_{\rm jet}^{-3}$, and \Eqref{eq:reg_bubble} ($z_{\rm iso}<R_{\rm Bondi}$) predicts a scaling of $\dot{M}\propto V_{\rm jet}^{-2}$.
These scaling relations roughly match what we see in \fref{fig:mdot_reg}. We plot the scaling relations fit to all of the runs with velocities $\gtrsim 10^4 {\rm km\,  s}^{-1}$ (more elongated cocoon), and with velocities $\lesssim 10^4 {\rm km\,  s}^{-1}$ (more bubble-like). As expected, the fits for the more elongated cocoon predict a steeper jet velocity dependence than the isotropic bubble case. It is also slightly steeper than what we predict from our simple model, but we re-emphasize that we are fitting a line to a small number of points and this result should be seen as a rough estimate.

When we vary the gas temperature (center-right panel), we find a scaling relation $\dot{M}_{\rm jet}\sim T_\infty^{-1.2}$. This is qualitatively consistent but a bit steeper than our model (\eeqref{eq:reg_bubble} and \eeqref{eq:reg_cocoon} with $\xi\sim0$), which implies a scaling to $V_{\rm jet}$ with a power-law index of \ksu{0 to -0.18 (cocoon) or -0.5 to -0.7 (bubble).  
} 
 
Finally, when we vary the gas density (bottom-left panel), the cocoon gas density depends on the background gas density as $n_c\propto n_\infty^\delta$ with $\delta\lesssim 0.9$. Therefore \Eqref{eq:reg_bubble} and \Eqref{eq:reg_cocoon} predict $\dot{M}_{\rm jet}\propto n_\infty^\alpha$ with $\alpha\gtrsim 1.1$. We plot the scaling relations fit to the runs with density $\gtrsim 10^4 {\rm km\, s}^{-1}$ (more elongated cocoon), and with density $\lesssim 10^4$ ${\rm km\, s}^{-1}$ (more bubble-like). 
The first scenario has a similar scaling relation to our toy model. The latter case results in a somewhat steeper slope, which qualitatively agrees with our toy model but is steeper than predicted.

\subsection{The growth of the black hole}\label{s:growth}

We indicate the mean \zh{time-averaged} black hole accretion rate of each model in \tref{tab:run}. The runs with the highest accretion rate are, unsurprisingly, the runs with the lowest feedback mass fraction ($\eta_{\rm m, fb}=0.005$), the lowest relative specific energy (that is the lowest jet velocity, $V_{\rm jet}=3000 {\rm \,km\, s}^{-1}$), the highest background density ($n_\infty=10^6 {\rm cm}^{-3}$), \ksu{and the lowest temperature ($T_\infty=10^3 K$).  They can all accrete at super-Eddington rates at their peak, reaching an accretion rate of $(10^{-6}-10^{-5}) {\rm M_\odot} {\rm yr}^{-1}$ or (0.4-6) $\dot{M}_{\rm Edd}$ on average, where the reference Eddington accretion rate relates to the Eddington luminosity as $\dot{M}_{\rm Edd}\equiv L_{\rm Edd}/0.1 c^2$ (although we remind readers that we are not treating radiative feedback in this work). In our surveyed parameter space, the presence of jet feedback suppresses the accretion rate below the Bondi rate by factors ranging from $\sim2\times10^{-4}$ up to 0.7. Note that there is a strong time variability of the black hole accretion rate and the resulting jet fluxes (see \fref{fig:flux_v}).} We only run the simulations for $<10^5$ yr, so none of the black holes grow significantly during the short periods covered by the simulations. Nevertheless, these results indicate that, for at least some of our model parameters, the black hole could grow to very large masses in cosmologically short times if it continues to be surrounded by high density gas.

\begin{figure}
    \centering
    \includegraphics[width=8cm]{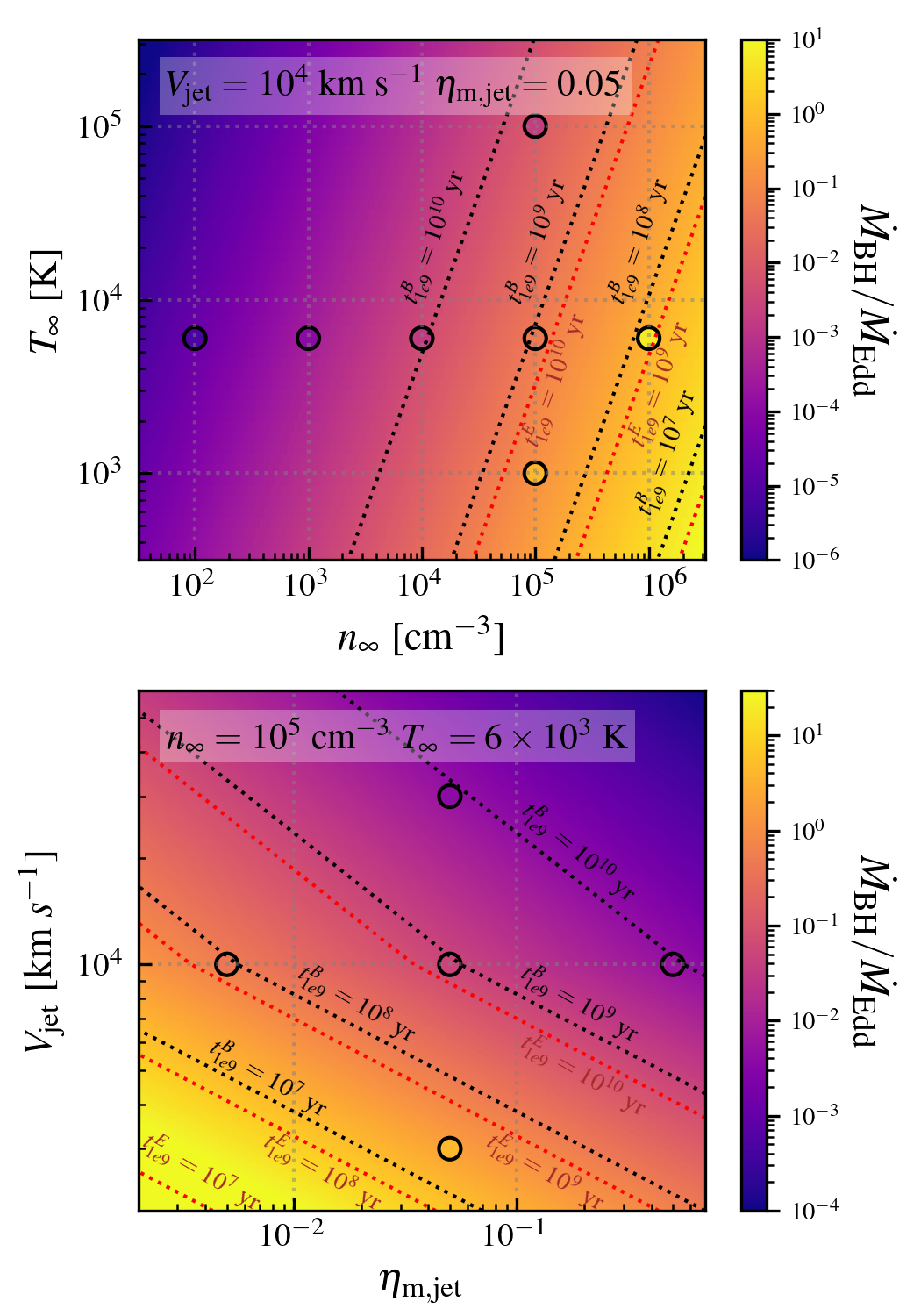}
    \caption{ \ksuu{The predicted $\dot{M}_{\rm BH}/\dot{M}_{\rm Edd}$ from the scaling of our toy model, assuming a normalization to the fiducial runs. Runs with low background gas temperature ($T_\infty$), high background gas density ($n_\infty$), low jet velocity ($V_{\rm jet}$), or low feedback mass fraction ($\eta_{\rm m, jet}$) result in super-Eddington accretion.
    The results from the simulations are shown as circles, coloured with the measured value. They show a qualitative agreement with the toy model. The dashed lines show the parameters for which the estimated time required for a 100 ${\rm M_\odot}$ black hole to grow to $10^9 {\rm M_\odot}$ ($t_{1e9}$) is $10^7$, $10^8$, $10^9$, and $10^{10}$ years. The calculation of $t^B_{1e9}$ assumes $\dot{M}_{\rm BH}\propto M_{\rm BH}^2$ (black, fixed fraction of Bondi rate with fixed background gas properties) throughout the evolution. The calculation of $t^E_{1e9}$ adopts \zh{a less optimistic extrapolation of the BH accretion rate} $\dot{M}_{\rm BH}\propto M_{\rm BH}$ (red, fixed fraction of the Eddington rate) throughout the evolution. }}
    \label{fig:mdot_map}
\end{figure}

\ksuu{We can express the ratio $\dot{M}_{\rm BH}/\dot{M}_{\rm Edd}$ using the scalings predicted by our toy model, normalized to the fiducial parameter choices, as
\begin{align}
\frac{\dot{M}_{\rm BH}}{\dot{M}_{\rm Edd}}\sim& 0.05 \left(\frac{\eta_{\rm m, BH}}{0.05}\right)^{-1}\left(\frac{V_{\rm jet}}{10^4 {\rm km/s}}\right)^{-2\,\, {\rm to}\, -3}\notag\\
&\left(\frac{n_\infty}{10^5 {\rm cm}^{-3}}\right)^{1.1}\left(\frac{T_\infty}{10^4 {\rm K}}\right)^{-0.6},
\end{align}
where the exponent of $V_{\rm jet}$ ranges from 2 (for $r_{\rm iso}<r_{\rm Bondi}$) to 3 (for $r_{\rm iso}>r_{\rm Bondi}$).
Assuming the separation of the two cases is roughly at $V_{\rm jet}\sim10^4 {\rm km s}^{-1}$, we plot the estimated BH accretion rate in \fref{fig:mdot_map}.  On top of the toy-model prediction, we indicate the results from our runs with circles colored with their measured values, and they show a qualitative agreement. We also show a set of dashed lines showing the parameters for which the estimated time needed for a 100 ${\rm M_\odot}$ black hole to grow to $10^9 {\rm M_\odot}$ ($t_{1e9}$) is $10^7$, $10^8$, $10^9$, and $10^{10}$ years. 
\zh{Since we only performed simulations for a single BH mass, $100~{\rm M_\odot}$, in this study, this requires extrapolating the time-averaged accretion rates to higher BH masses.}
The calculation of $t_{1e9}$ assumes $\dot{M}_{\rm BH}\propto M_{\rm BH}^2$ (corresponding to a fixed fraction of the Bondi rate, with fixed background gas properties) throughout the evolution. This is motivated by the $M_{\rm BH}$ dependence of $\dot{M}_{\rm jet}$ predicted in our toy model  (see \Eqref{eq:reg_cocoon} and \Eqref{eq:reg_bubble}) but will be left for future study to verify with simulations with different black hole masses. \zh{For a less optimistic estimate}, the calculation of $t^E_{1e9}$ instead assumes $\dot{M}_{\rm BH}\propto M_{\rm BH}$ (corresponding to a fixed fraction of the Eddington rate) throughout the evolution. Given the assumptions above and the estimated accretion rate of each case at $\dot{M}_{\rm BH}=100 {\rm M_\odot}$, part of the parameter space can have a 100 ${\rm M_\odot}$ black hole growing to a $10^9 {\rm M_\odot}$ supermassive black hole at high redshift. We emphasize that these are crude estimations. The underlying assumptions of a fixed fraction of Bondi accretion and the constant background gas properties are subject to verification in future work.}

\subsection{Jet duty cycle}\label{s:duty}

\begin{figure}
    \centering
    \includegraphics[width=8cm]{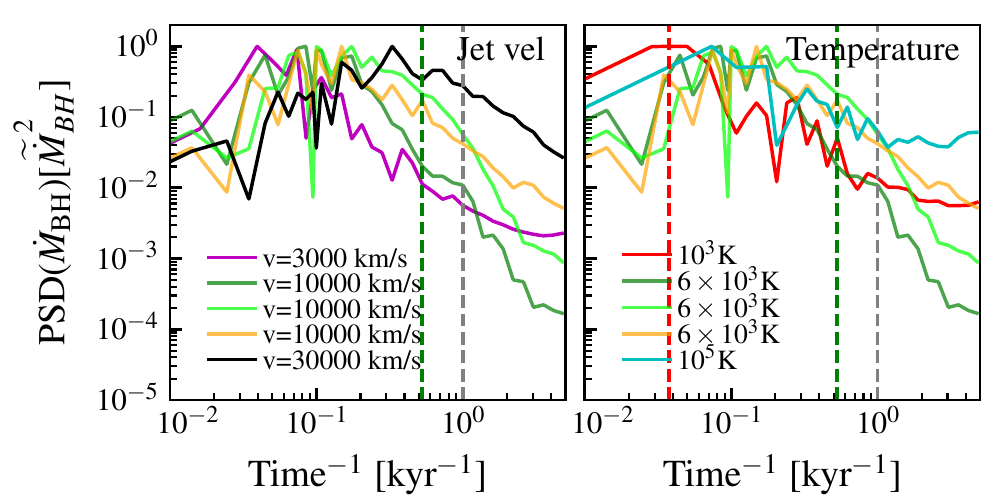}
    \caption{\ksu{The normalized (divided by the maximum value) power spectrum of the black hole accretion history. The green and red vertical lines are the free-fall time at the Bondi radius for $6\times10^3$ K and $\times10^3$ K gas. The corresponding value for $\times10^5$ K is outside the plotted range (right end). The gray line is the viscous time scale of the $\alpha$-disk.} The lower jet velocity run results in a more elongated cocoon, which reaches a larger distance, and has longer-term variability. The run with cooler background gas also has a larger Bondi radius, so the black hole accretion rate also has longer-term variability.}
    \label{fig:mdot_duty}
\end{figure}

Besides regulating the black hole accretion rate and jet mass flux, the various jet models and background gas properties also affect the feedback cycle period. A run with a more elongated jet cocoon that propagates to a larger distance will result in longer-term variability in the accretion rate. This can be seen in the left panel of \fref{fig:mdot_duty}, where we quantify the normalized (i.e. divided by its maximum value) power spectrum of \zh{the BH accretion rate} in the runs with different velocities. We clearly see that the slower the jet velocity, the more elongated the jet cocoon becomes and the more the power spectrum shifts to longer periods (lower frequencies). The reason for this behavior is simply that, when the jet reaches a larger distance, the time scale of the regulation (i.e., the free fall time) becomes longer.

Similarly, changing the gas temperature also impacts the distance that the cocoon reaches. 
In the right panel of \fref{fig:mdot_duty}, we see that the higher temperature run, which has the smaller Bondi radius, has a power spectrum shifted to shorter periods. 

\section{Discussion}\label{S:discussion}

\subsection{Comparison with previous works}\label{S:previous}

\zh{\citet{2019MNRAS.486.3892R} used the adaptive mesh refinement (AMR) hydrodynamics code \texttt{ENZO} to  investigate feedback from bipolar jets, expected to be produced during super-Eddington accretion episodes, focusing on how the jet feedback impacts BH growth.  They found that the jets periodically evacuate the central $\sim 0.1$pc region, and accretion then resumes after a free-fall time.  Overall, we here find a similar behaviour, although there are several differences between our setups and our results.    \citet{2019MNRAS.486.3892R} utilize a cosmological simulation, and adopt an initial seed BH mass of $16,000~{\rm M_\odot}$, with an initial accretion rate of $\sim10^{-2}~{\rm M_\odot}\, {\rm yr}^{-1}$, and find that the time-averaged accretion rate always stays below the Eddington value.  Also, once the gas is heated by the jet, they do not resolve the Bondi radius, and adopt a modified Bondi accretion rate.  By comparison, we here use an initially uniform and static cloud, and examine a $>100$ times lower BH mass, $>100$ times lower accretion rate, and a $>100$ times higher spatial resolution, such that the Bondi radius remains resolved at all times.  We find that super-Eddington accretion is possible, which may be explained by the different parameter choices and/or differences in the details in the shapes of the jet-driven cocoons.  Another difference between our studies is that we examine the cocoon evolution in greater detail and offer a physical interpretation of its shape and size, as a function of jet and background gas parameters.}

\cite{2011ApJ...739....2P} studied a similar problem regarding accretion onto a low-mass black hole, but with radiative feedback instead of the mechanical jet feedback explored here. They obtain a scaling of the time-averaged black hole accretion rate with background gas properties ($\dot{M}\propto T_\infty^{5/2}$ for $n_\infty\geq10^5 {\rm cm}^{-3}$ and $\dot{M}\propto T_\infty^{5/2} n_\infty^{1/2}$ for $n_\infty<10^5 {\rm cm}^{-3}$), which are different from ours. The primary reason is that the radiative feedback they implemented inflates a roughly constant temperature bubble around the black hole, which is in pressure equilibrium with the surrounding cold gas. This is very different from the cocoon we see inflated by jet feedback, where the cocoon has a more complicated shape as discussed in \sref{S:toy}. \cite{2011ApJ...739....2P}  also found a feedback cycle with a well-defined period, while we have much more complicated cycles. \ksu{This can arise from the more anisotropic turbulent gas distribution due to the jetted feedback or the more complex geometry of the outflows and accretion.} The fact that we are using 3D simulations, while \cite{2011ApJ...739....2P} used 1D and 2D simulations, could also contribute to the difference.
\zh{Overall, the time-averaged accretion rate in \cite{2011ApJ...739....2P} was found to always remain below the Eddington rate, whereas we here find super-Eddington accretion in many cases.  This suggests that jet feedback may be a lesser obstacle to BH growth than radiative feedback.}

\ksu{\cite{2020MNRAS.497..302T} also studied a similar problem with different black hole masses ($10$ and $10^5 {\rm M_\odot}$) and with wider AGN winds on top of radiative feedback. \zh{They focused on the 'hyper-Eddington' regime, such that the Bondi rate exceeds the Eddington rate by several orders of magnitude. In practice, they considered either a much higher BH mass or a much higher background gas density than in our study.} They showed that \zh{under these conditions}, the resulting accretion rate can remain close to the Bondi accretion rate and reach \zh{the prescribed hyper}-Eddington values after around a dynamical time, when the radiative feedback becomes less important. They also found that the accretion rate is insensitive to the feedback mass fraction of the mechanical feedback. These latter findings are qualitatively similar to what we see in our simulations with the AGN jet, despite different initial velocity and initial open-angle. They also see momentum conserving wind propagation (constant velocity) all the way to beyond Bondi radius, qualitatively similar to what we see in our lower velocity jets, which is also expected in our toy model.
Given the difference in the feedback form, black hole mass, and run time, \zh{and background Bondi accretion rate}, more quantitative comparisons would be difficult.}

It is worth also comparing our results with the jets in larger-scale galaxy simulations, e.g. those presented in \cite{2021MNRAS.507..175S}. A similar qualitative result was found in those galaxy scale studies, namely that heavier jets result in much narrower jet cocoons which propagate much farther. The toy model describing the jet propagation presented in \sref{S:toy} also works on galaxy scales with a much more massive black hole and lower gas density. One important difference is the relative strength of radiative cooling, which operates more rapidly in the current simulations. Here, we find significant cooling at the contact discontinuity between the shocked jet material and the shocked ambient medium, whereas cooling on the cluster and galaxy scale is slower and occurs mostly in gas which is not shock-heated. 

\zh{\citet{2022arXiv220108766M} also examined the impact of jet feedback on BH growth on larger ($\sim$kpc scales), in a $10^{11}~{\rm M_\odot}$ dark matter halo and found that mildly super-Eddington accretion is possible. They found that weaker super-Eddington
jets allow for more rapid BH growth through more frequent super-Eddington episodes, and also that 
weaker jet feedback efficiency leads to larger BH masses,  which are consistent with our findings.}

\subsection{Connection to other scales}\label{s:scale}

Part of the motivation of this work, where we perform intermediate-scale simulations, is to provide insight in connecting galaxy scale simulations and GRMHD simulations that can resolve the accretion disk. Depending on the galaxy size and the numerical method, the finest resolution of the former case is at best $\sim 0.1$ pc, and generally much lower \citep[e.g.,][]{2016ApJ...827L..23W,2018MNRAS.480.1666S,2019MNRAS.490.4447W,2022arXiv220108766M}. The outer boundary of the latter case is at most $\sim 1000r_g$ ($r_g$ is the gravitational radius; e.g. \citealt{2022arXiv220208281L}), which is roughly $10^{-8} (M_{\rm BH}/100 {\rm M_\odot})$ pc, implying a $\gtrsim 7$ order of magnitude gap for the black hole mass we model here. Our simulation, with its outer boundary at roughly $0.2$ pc and a maximum resolution of $\sim 10^{-4}$ pc fits between these scales, although we note that we are still far from $\sim 1000r_g$.

Unless using GRMHD simulations, which resolve the gravitational radius, AGN jets are not self-consistently launched but are implemented instead with sub-grid prescriptions. Effectively these sub-grid ``jet models'' attempt to inject the fluxes of a ''cocoon'' inflated by a jet launched on an even smaller scale. Therefore, even identical jet energy, momentum, and mass fluxes can produce different physical behaviour when launched on different scales. This work provides a framework for coarse-graining jet models launched on a smaller scale to the resolution scale of galaxy simulations. The toy model described in \sref{S:toy} and verified in our simulations describes how the cocoon energy, momentum, and mass flux should scale as a function of radius. The scalings can be incorporated into simulations on different scales for the same sub-grid jet model.

\ksu{Effectively, given a certain estimated density ($n_\infty$) and temperature ($T_\infty$) around a black hole, and a jet model ($V_{\rm jet}$) on a small scale, $r_{\rm small}$, 
\Eqref{eq:criteria} 
can roughly determine whether $z_{\rm iso}$ is larger or smaller than $R_{\rm Bondi}$. Depending on which side it falls, the resulting time-averaged $\dot{M}_{\rm jet}$ can be estimated through \Eqref{eq:reg_cocoon} or \Eqref{eq:reg_bubble}. Note that we also need the $\zeta$, $\xi$, and $\delta$ values from the fit results in \fref{fig:phase_fit}. Assuming we want to find the effective cocoon property at a given larger radius, $r_{\rm large}$, the cocoon expansion either follows \Eqref{eq:V_r_time} and \Eqref{eq:V_z_time} (if $r_{\rm large}<r_{\rm iso}$) or \Eqref{eq:iso_r_time} (if  $r_{\rm large}>r_{\rm iso}$). 
Therefore, with the values $n_\infty$, $T_\infty$, $V_{\rm jet}$, $\zeta$, $\xi$, $\delta$, and $\dot{M}$, we find the corresponding cocoon expansion velocity at a specific radius $r_{\rm large}$, which can be used as an ``effective'' coarse-grained jet model at that scale. The aforementioned implementation should, of course, be explicitly tested in galaxy-scale simulations. Indeed, besides the effective velocities, there is also the complexity of an ``effective'' jet model, including the temperature, time variability, gas cooling, and the exact sampling of the velocity distribution while launching the feedback. We leave a thorough investigation of these issues, and the construction of a full sub-grid jet recipe, to future work.}

\subsection{Limitations of this work and future prospects}\label{s:future}

We emphasize that we have deliberately considered an idealized setup, with an initially static cloud with a uniform density and temperature. In reality, the gas surrounding the black hole could be highly turbulent with a non-zero net angular momentum. 
We also consider only one black hole mass \citep[see e.g.,][for similar studies with larger black holes. ]{2019MNRAS.486.3892R,2020MNRAS.497..302T,2022arXiv220108766M}. Moreover, this work does not include magnetic fields, conduction, viscosity, and other plasma physics, which may be important on these scales. 

For the feedback itself, we only include jet feedback in this work for simplicity, ignoring any radiative feedback \citep[e.g.,][]{2011ApJ...739....2P,2019MNRAS.486.3892R,2020MNRAS.497..302T}, which may also play an important role in the black hole's neighborhood. Due to the limitations of non-relativistic hydrodynamics, we also limit the jet velocity to $\lesssim 30000\, {\rm km\,  s}^{-1}$. This should be reasonable at the jet launching scale of our simulations but might not cover the whole possible parameter space in more extreme circumstances. We also did not explore models with wider opening-angle AGN winds \citep[e.g.,][]{2020MNRAS.497..302T}. Cosmic rays might be another critical aspect of AGN feedback as well \citep{2020MNRAS.491.1190S,2021MNRAS.507..175S,2022arXiv220306201W}, but are not included here. We will explore these aspects in future work.

\section{Conclusion}\label{S:conclusions}

In this work, we utilized high-resolution hydrodynamic simulations of 0.4-1.6 pc boxes with uniform initial density and temperature to study jet propagation and its effect on black hole accretion onto a 100 ${\rm M_\odot}$ black hole in low metallicity dense gas. We found that the isotropic component of the cocoon momentum flux regulates the black hole accretion and the mass, momentum, and energy flux from the AGN jet. We summarize our major conclusions as follows:

\begin{itemize}
    \item After a jet is launched, it inflates a jet cocoon filled with a hot reverse shock-heated turbulent gas and a much cooler gas phase at the mixing layer with the surrounding gas.
    \item At launch, a jet cocoon will propagate, conserving the momentum in the jet direction while continuously broadening itself through \zh{thermal pressure} in the lateral directions. Eventually, the cocoon expands laterally and the propagation in the jet direction slows down. \ksu{If the jet cocoon propagates to a sufficiently large radius, it eventually \zh{evolves into a quasi-spherical bubble}. After that, the cocoon propagates isotropically in an energy-driven scenario.}
    \item Depending on the jet and background gas properties, the inflated cocoon either isotropizes beyond the Bondi radius (retaining an elongated shape), or inside the Bondi radius (becoming spherical bubble-like).
    \item In either case, the isotropic component 
    of the cocoon momentum flux \ksu{(roughly twice the lateral momentum flux if the cocoon is elongated)} on average matches the inflow momentum flux at the Bondi radius, assuming a Bondi-accretion scenario. This, in turn, regulates the black hole accretion.
    \item We presented a toy model based on this picture which results in a scaling of the black hole accretion rate that roughly matches the rate found in the simulations.
    \item The lower the jet velocity and the higher the background gas density, the more elongated the jet cocoon.
    \item In addition to the average black hole accretion rate and jet mass flux, the different jet model and background gas properties also affect the accretion history variability. A jet model that produces an elongated cocoon propagates to a larger distance and produces longer-timescale variability, while smaller and more spherical bubble-like cocoons produce shorter-timescale variability. Higher $T_\infty$ (smaller $R_{\rm Bondi}$) also leads to more short-timescale variability.
    \item  \ksu{The runs with the highest accretion rates are those with the lowest feedback mass fraction ($\eta_{\rm m, fb}=0.005$), the lowest specific energy or jet velocity ($V_{\rm jet}=3000 {\rm \,km\, s}^{-1}$), the highest density ($n_\infty=10^6 {\rm cm}^{-3}$), or the lowest temperature ($T_\infty=10^3 K$). They, on average, have super Eddington accretion, $\dot{M}_{\rm BH}\sim 0.4-6\dot{M}_{\rm Edd}$. In our surveyed parameter space, the presence of AGN jets suppresses the Bondi accretion rate by factors from $\sim2\times10^{-4}$ to $0.6$.}
\end{itemize}

In summary, this work shows how different jet models (and background gas properties) result in different cocoon properties and accretion rates. \zh{Our results suggest that at least initially, stellar-mass black holes in so-called 'atomic cooling haloes' may be able to grow at rates well above the Eddington rate.} 
Our study \zh{also suggests a prescription to link} simulations on different scales (\sref{s:scale}). Many caveats and unanswered questions remain (see \sref{s:future}) to be explored in future work.

\vspace{-0.2cm}
\acknowledgments
Numerical calculations were run on the Flatiron Institute cluster ``popeye'' and ``rusty'' and allocations TG-PHY220027 and  TG-PHY220047 granted by the Engineering Discovery Environment (XSEDE)
supported by the NSF. KS acknowledges support from Simons Foundation. GLB acknowledges support from the NSF (OAC-1835509, AST-2108470), a NASA TCAN award, and the Simons Foundation.
\zh{ZH acknowledges support from NSF grant AST-2006176}.  RSS and CCH were supported by the Simons Foundation through the Flatiron Institute. CAFG was supported by NSF through grants AST-1715216, AST-2108230,  and CAREER award AST-1652522; by NASA through grants 17-ATP17-006 7 and 21-ATP21-0036; by STScI through grants HST-AR-16124.001-A and HST-GO-16730.016-A; by CXO through grant TM2-23005X; and by the Research Corporation for Science Advancement through a Cottrell Scholar Award.
\vspace{0.3cm}
\section*{Data Availability statement}
The data supporting the plots within this article are available on reasonable request to the corresponding author. A public version of the GIZMO code is available at \href{http://www.tapir.caltech.edu/~phopkins/Site/GIZMO.html}{\textit{http://www.tapir.caltech.edu/$\sim$phopkins/Site/GIZMO.html}}.

\bibliographystyle{mnras}
\bibliography{mybibs}

\appendix
\normalsize
\section{Resolution Study and the variations of accretion models}
\label{a:res}
\begin{figure}\label{fig:res}
    \centering
    \includegraphics[width=8cm]{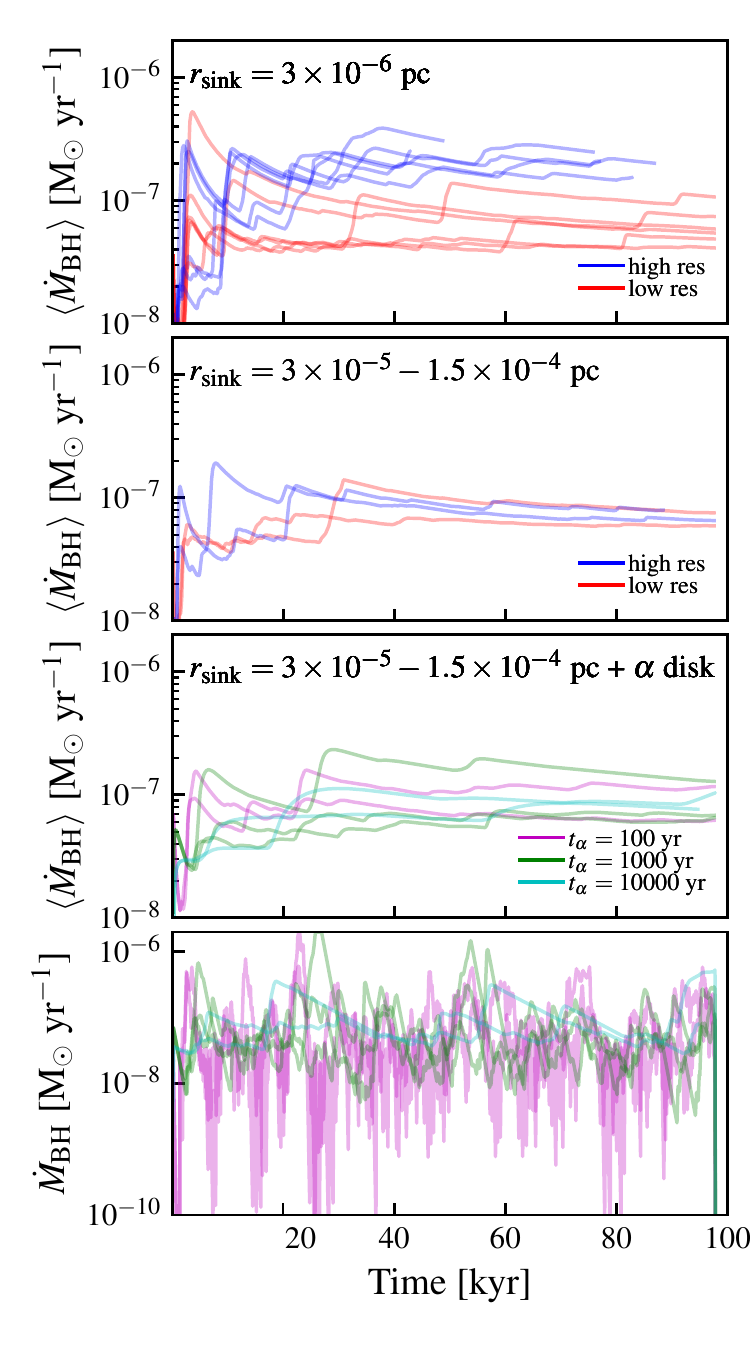}
    \caption{The effects of different choices of sink radius and alpha disk models on black hole accretion rate $\dot{M}_{\rm BH}$ under different resolutions. The first three rows show the time-averaged accretion rates for a range of sink radii, resolutions, and alpha disk parameters. To explore stochastic variations, we run simulations are run with different variations of the random component (different lines in the same color). With the smallest sink radius (0.003 mpc), the random number variations result in a factor of 2-3 span in the final results, indicating that stochastic effects are significant. The higher resolution runs also result in a factor of 2-3 higher $\dot{M}_{\rm BH}$. The runs with a larger sink radius have slightly better convergence and smaller stochastic effects ($\lesssim 2$). The model with alpha disks and different viscous time scales also have accretion rates with differences within a factor of 2, well within the stochastic range. The bottom panel shows the real-time $\dot{M}_{\rm BH}$ of the runs with different viscous time scales. A shorter viscous time scale results in shorter-term variations.}
    \label{fig:res_acc}
\end{figure}

\begin{table}
\begin{center}
 \caption{Physics variations (run at highest resolution) explored in this appendix}
 \label{tab:res}
\resizebox{8cm}{!}{%
\begin{tabular}{c|cccc|cc}
\hline
\hline
&\multicolumn{4}{c|}{} &\multicolumn{2}{c}{Accretion model}  \\
\hline
Model           & $\Delta T$ & Box size & $m^{\rm max}_{\rm g}$ & $m_{\rm jet}$ & $r_{\rm sink}$ & $t_\alpha$ \\
                & kyr        & pc       & ${\rm M_\odot}$             & ${\rm M_\odot}$   & $10^{-3}$pc & kyr   \\ 
\hline
\multicolumn{5}{l|}{\bf \underline{\hlt{$\,\,\,r_{\rm sink}=3\times10^{-6}$ pc\,\,\,}}}\\
high res & 40-80 & 0.4 & 1.7e-7 & 3e-8 &0.003 & No \\
low res & 100 & 0.4 & 1.4e-6 & 1e-7 &0.003 & No \\
\hline
\multicolumn{5}{l|}{\bf \underline{\hlt{$\,\,\,r_{\rm sink}=3\times10^{-5}-1.5\times10^{-4}$ pc\,\,\,}}}\\
high res & 100 & 0.4 & 1.7e-7 & 3e-8 &0.03 -0.15 & No \\
low res & 100 & 0.4 & 1.4e-6 & 1e-7 &0.03 -0.15 & No \\
\hline
\multicolumn{5}{l|}{\bf \underline{\hlt{$\,\,\,r_{\rm sink}=3\times10^{-5}-1.5\times10^{-4}$ pc + $\alpha$ disk\,\,\,}}}\\
100 yr & 100 & 0.4 & 1.4e-6 & 1e-7 &0.03-0.15 & 0.1 \\
1000 yr & 100 & 0.4 & 1.4e-6 & 1e-7 &0.03-0.15 & 1 \\
10000 yr & 100 & 0.4 & 1.4e-6 & 1e-7 &0.03-0.15 & 10 \\
\hline 
\hline
\end{tabular}
}
\end{center}
\begin{flushleft}
This is a partial list of simulations that explore resolution and numerical parameter choice. All simulations are run with ($\zeta_{\rm m, fb}=0.05$, $V_{\rm jet}=10^4 {\rm km\,  s}^{-1}$, $n_\infty=10^5{\rm cm}^{-3}$, and $T_\infty=10^4 K$ ).
Columns list: 
(1) Model name:  The naming of each model starts with the feedback mass fraction, followed by the jet velocity in km s$^{-1}$ for kinetic jet or jet temperature in K for thermally dominant jets. The final two numbers label the background gas density in cm$^{-3}$ and temperature in K. 
(2) $\Delta T$: Simulation duration. 
(3) Box size of the simulation.
(4) $m^{\rm max}_{\rm g}$: The highest mass resolution.
(5) $m^{\rm max}_{\rm jet}$: The mass resolution of the spawned jet particles.
(6) $r_{\rm sink}$: Sink radius in $10^{-3}$ pc.
(7) $t_\alpha$: Viscous time scale for alpha disk in kyr.
\end{flushleft}
\end{table}

\fref{fig:res} summarizes the effects of different choices of sink radius and alpha disk model on black hole accretion rates $\dot{M}_{\rm BH}$ under different resolutions. All the runs match our fiducial parameter choice ($\zeta_{\rm m, fb}=0.05$, $V_{\rm jet}=10^4 {\rm km\,  s}^{-1}$, $n_\infty=10^5{\rm cm}^{-3}$, and $T_\infty=10^4 K$ ). The first 3 rows shows the averaged value to the point at the specific time of the simulation. 
Most simulations are run with different variations of the random component to quantify the stochastic effect (different lines in the same color). A list of different simulations is summarized in \tref{tab:res}.

With the smallest sink radius ($3\times10^{-6}$ pc), the stochastic effects result in a factor of 2-3 span in the final results indicating a substantial stochastic effect. The higher resolution runs also result in a factor of 2-3 higher $\dot{M}_{\rm BH}$. \ksu{The small sink radius also leads to the occasional formation of a disky structure right around the black hole at high resolution, which partially contributes to the more significant resolution dependence. Given that we do not have the proper resolution and physics to model the accretion disk explicitly, we shift to a larger sink radius and put in a subgrid $\alpha$-disk model as described in the main paper.}

The runs with a larger sink radius ($3\times10^{-5}-1.5\times10^{-4}$ pc) \footnote{\ksu{The sink radius is set to be a radius from the black hole enclosing 96 ``weighted'' neighborhood gas particles but capped to be within  ($3\times10^{-5}-1.5\times10^{-4}$ pc).}} \ksu{have a slightly smaller dependence on resolution and smaller stochastic effects (everything within $\lesssim 2$), partially due to the suppression of an artificial disky structure at very small radius.} This level of difference (even the small sink radius runs) is smaller than the difference caused by most of the physics variations \fref{fig:flux_v} and \fref{fig:mdot_reg}. {\bf In our production run, we adopt the larger sink radius ($3\times10^{-5}-1.5\times10^{-4}$ pc).} Given the \ksu{smaller resolution dependence} with this sink radius, we try to match the lower resolution for most of our physical variations for lower computational cost.

The models with alpha disk and different viscous time scales also result in differences within a factor of two, within the stochastic range, and roughly have the same accretion rates as the runs without an alpha disk. The final row of \fref{fig:res} shows the real-time $\dot{M}_{\rm BH}$ of the runs with different viscous time scales. Shorter time scale results in a shorter-term variation. {\bf We adopt $t_\alpha=1000$ yr in our productive runs according to an estimate of the viscous time scale at the sink radius we choose (see \sref{S:methods})}.
\label{lastpage}

\end{document}